Roux J, Privman E, Moretti S, Daub JT, Robinson-Rechavi M, Keller L. 2014. Patterns of positive selection in seven ant genomes. Mol Biol Evol.

Article

# Patterns of positive selection in seven ant genomes

Julien Roux[1,2,5], Eyal Privman[1,2,6], Sébastien Moretti[1,2,3], Josephine T. Daub[1,2,4], Marc Robinson-Rechavi[1,2], Laurent Keller[1]

[1] Department of Ecology and Evolution, University of Lausanne, 1015 Lausanne, Switzerland

[2] SIB Swiss Institute of Bioinformatics, 1015 Lausanne, Switzerland

[3] Vital-IT group, SIB Swiss Institute of Bioinformatics, 1015 Lausanne, Switzerland

[4] CMPG, Institute of Ecology and Evolution, University of Bern, Baltzerstrasse 6, 3012 Bern, Switzerland

[5] Present address: Department of Human Genetics, University of Chicago, Chicago, IL 60637, USA

[6] Present address: Department of Evolutionary and Environmental Biology, University of Haifa, Israel

Corresponding author: Julien Roux (Julien.Roux@unil.ch)

## Short title

Positive selection in ant genomes

## Authors contributions

JR, EP and LK designed the study; JR and EP analyzed data; JTD, SM and MRR contributed code or programs; JR and LK wrote the manuscript with input from MRR.

## Abstract

The evolution of ants is marked by remarkable adaptations that allowed the development of very complex social systems. To identify how ant-specific adaptations are associated with patterns of molecular evolution, we searched for signs of positive selection on amino-acid changes in proteins. We identified 24 functional categories of genes which were enriched for positively selected genes in the ant lineage. We also reanalyzed genome-wide datasets in bees and flies with the same methodology, to check whether positive selection was specific to ants or also present in other insects. Notably, genes implicated in immunity were enriched for positively selected genes in the three lineages, ruling out the hypothesis that the evolution of hygienic behaviors in social insects caused a major relaxation of selective pressure on immune genes. Our scan also indicated that genes implicated in neurogenesis and olfaction started to undergo increased positive selection before the evolution of sociality in Hymenoptera. Finally, the comparison between these three lineages allowed us to pinpoint molecular evolution patterns that were specific to the ant lineage. In particular, there was ant-specific recurrent positive selection on genes with mitochondrial functions, suggesting that mitochondrial activity was improved during the evolution of this lineage. This might have been an important step toward the evolution of extreme lifespan that is a hallmark of ants.

# Introduction

Ants constitute an extremely successful lineage of animals which has colonized virtually all ecosystems on Earth (Hölldobler, Wilson 1990). The pivotal feature at the basis of this ecological success is their highly social system with a reproductive division of labor, where one or a few queens specialize in reproduction, while workers conduct all the colony tasks such as brood care, nest maintenance and food collection. In this paper we take advantage of the recent availability of seven sequenced ant genomes (Bonasio et al. 2010; Nygaard et al. 2011; Smith et al. 2011a; Smith et al. 2011b; Suen et al. 2011; Wurm et al. 2011) to perform a genome-wide scan for positive selection on amino-acid changes in protein coding genes during the evolution of the ant lineage. We addressed three main questions.

First, we compared the amount of positive selection in functional categories of genes. Previous large-scale scans for positive selection in animals indicated that positive selection predominantly affects certain types of genes, such as those involved in evolutionary arms races, sexual selection or conflicts with pathogens (Bakewell, Shi, Zhang 2007; Drosophila 12 Genomes Consortium 2007; Kosiol et al. 2008; Vamathevan et al. 2008; Oliver et al. 2010; George et al. 2011; Woodard et al. 2011). Such genes experienced positive selection events recurrently on broad evolutionary time scales, and it is likely that they contribute to a fraction of the positive selection events that occurred in the ant lineage. To identify these genes, we reasoned that they likely also were under positive selection in other insect lineages. A systematic comparison of the targets of positive selection from published studies in insects is not straightforward because genome-wide scans for positive selection were often performed with different methods in different lineages. For example, a positive selection scan on 12 Drosophila species (all solitary)(Drosophila 12 Genomes Consortium 2007) used the site test of Codeml (Yang et al. 2000), which is aimed at detecting recurrent positive selection events affecting particular sites of a protein, while a scan on 10 bee species (including solitary, primitively social and highly social species)(Woodard et al. 2011) used the branch test (Yang 1998), which tends to detect positive selection events affecting a large number of sites of a protein but during a limited period of time. To perform a robust comparison of the genes that were under positive selection in ants and other insects, we conducted similar scans for positive selection in ants and the flies and bees outgroups. An example of genes expected to be repeatedly under positive selection in insects are genes involved in defense and immunity (Drosophila 12 Genomes Consortium 2007; Bulmer 2010). On the basis of the observed smaller set of immunity genes in the honeybee compared to *D. melanogaster,* it has been suggested that selective pressure on these genes might have been relaxed in social insects, perhaps because they have social hygienic behaviors (Honeybee Genome Sequencing Consortium. 2006; Smith et al. 2008; Viljakainen et al. 2009; Smith et al. 2011a; Suen et al. 2011; Harpur, Zayed 2013). However, the addition of several newly sequenced insect genomes revealed that the important gene complement in fruit fly is a derived character (Werren et al. 2010; Fischman, Woodard, Robinson 2011; Smith et al. 2011b). We used our datasets to test whether there was evidence for weaker positive selection on immunity genes in ants and bees compared to flies.

Next, we aimed at detecting sets of genes involved in functions likely to reflect ant-specific adaptations. We focused on three main adaptations. The first relates to the wide range of coordinated collective behaviors associated with division of labor in ant societies. Complex cooperative behaviors occur among nestmates for tasks such as communal nest construction and defense, brood rearing, social hygienic behavior and collective foraging (Hölldobler, Wilson 1990). It has been suggested that the evolution of social interactions may be tracked down to molecular changes affecting nervous system development and function. In particular it may translate into increased rates of positive selection on nervous system-related genes, as documented in primitively social lineages of bees, which evolved social behaviors independently from ants (Fischman, Woodard, Robinson 2011; Woodard et al. 2011). Complex collective behaviors also require efficient communication systems, that are essentially mediated by chemical signaling in social insects. Ants identify nestmates from non-nestmates, as well as ants from other species, through their scent. Individuals also use various types of pheromones as alarm signals and to mark their trails and territories. It has therefore been suggested that genes involved in chemical signaling, notably pheromone production and perception, should experience increased positive selection in ants compared to solitary insects (Ingram, Oefner, Gordon 2005; Robertson, Wanner 2006; Bonasio et al. 2010; Smith et al. 2011a; Wurm et al. 2011; Kulmuni, Wurm, Pamilo 2013; Leboeuf, Benton, Keller 2013). A manually curated dataset of 873 olfactory receptor genes allowed us to conduct a test for increased positive selection on these genes in ants.

The second type of potential molecular adaptation relates to phenotypic plasticity among castes. Although queens and workers usually develop from totipotent eggs (Schwander et al. 2010), they display dramatic morphological and physiological differences. Queens are often larger, have wings and have much more highly developed ovaries than workers which often are sterile and lack a sperm storage organ (Hölldobler, Wilson 1990). In most species the differences between castes result from developmental differences induced by environmental factors rather than genetic differences (Abouheif, Wray 2002; Schwander et al. 2010; Penick, Prager, Liebig 2012; Rajakumar et al. 2012). We therefore investigated whether there was evidence for increased positive selection in genes and pathways potentially involved in developmental plasticity (Smith et al. 2008; Fischman, Woodard, Robinson 2011).

A third and interesting type of ant-specific adaptation relates to the extremely long lifespan of ant queens, which can live more than 20 years in some species (Keller, Genoud 1997; Jemielity et al. 2005). This corresponds to a 100-fold increase in lifespan compared to solitary insects. The variation in lifespan among castes is also remarkable, with queens living up to 10 times longer than workers and 500 times longer than males. So far, a limited number of molecular candidates have been identified to explain this pattern, mainly inspired from work in Drosophila (Jemielity et al. 2005; Keller, Jemielity 2006). We therefore investigated if there was evidence of positive selection on genes that have previously been associated with aging in model organisms. It is possible that positive selection acted on the same sets of genes in the bee lineage, where queens also live longer than other castes and than solitary insects, but such a signal should not be observed in short-lived species of the Drosophila lineage. To further assess the link between positive selection and aging, we investigated whether genes that experienced positive selection in ants were genes shown in *Drosophila melanogaster* to be differentially expressed between old and young individuals, and between oxygen-stressed and control individuals (Landis et al. 2004).

Finally, we investigated whether there was a difference in the level of positive selection between genes showing biased expression in queens, workers and males. The efficiency of natural selection acting on an advantageous mutation—and thus the probability of its long-term fixation—is proportional to its effect on fitness (Duret 2008). The fitness effects of mutations in genes that are expressed only in non-reproductive workers are indirect, so everything else being equal, selection should be less efficient at fixing them than mutations on genes expressed in queens and males. This could translate into lower levels of positive selection on genes expressed specifically in workers compared to males and queens (Linksvayer, Wade 2009; Hall, Goodisman 2012). We therefore analyzed previously published microarray data from the red fire ant *Solenopsis invicta* (Ometto et al. 2011), and compared the amount of positive selection between groups of genes varying in the level of caste-biased expression.

## Results

### *Pervasive positive selection detected in ants*

To detect positive selection episodes that acted on protein coding genes during the evolution of the ant lineage, the branch-site test of Codeml was run on 4,261 protein alignments of single-copy orthologs composed of four to seven ant and three to five outgroup species (see Materials and Methods). All branches that led to ant species in each gene family tree (including 2 hymenopteran and 13 ant branches, Figure 1) were successively tested for the presence of episodic positive selection. As many as 1,832 single-copy orthologs families (43%) displayed a signal of positive selection (at 10% FDR) in at least one of the branches tested (Table S1). In 91% of the significant alignments, at least one residue targeted by positive selection could be identified with a posterior probability greater than 0.9 (Bayes Empirical Bayes test, Figure 2)(Yang, Wong, Nielsen 2005). There was evidence for positive selection in at least one branch of the ant lineage for 830 (20%) of the genes analyzed. For 74% of them positive selection was specific to ants and not observed in the basal hymenopteran branches #7 and #8 (Figure 1). The 10 gene families with the most significant test values in the ant lineage are given in Table 1.

The proportion of positively selected genes varied significantly across the different branches tested ($\chi^2$ test $p<10^{-15}$; Table 2), similarly to previous analyses with experimental and simulated datasets. This likely results at least in part from lower power of the branch-site test in shorter branches (Anisimova, Yang 2007; Kosiol et al. 2008; Studer et al. 2008; Fletcher, Yang 2010; George et al. 2011; Gharib, Robinson-Rechavi 2013). Consistent with this view, there was a significant correlation in our dataset between the length of tested branches and the test score (log-likelihood ratio;

Spearman correlation $\rho=0.41$, $p<10^{-15}$). Additional analyses ruled out the hypothesis that false positives caused by either convergence problems of the test, selective constraints acting on synonymous sites, saturation of synonymous substitution rate $d_S$ or sequencing errors could be responsible for this pattern (Supplementary text).

Taken together, these results demonstrate that positive selection was common in the evolution of the ant genes. The proportion of significant genes was similar in magnitude in the outgroup dataset of 10 bees analyzed with the same methodology (20%; Table S2), but even higher in the outgroup dataset of 12 flies (36%; Table S3). This difference might reflect biological differences between the lineages, such as effective population size $N_E$, but also differences in the topology and branch lengths of the species trees, which influence the power to detect positive selection events in protein alignments (see Supplementary text).

To compare the amount of positive selection experienced by different functional categories of genes, we classified genes based on their Gene Ontology (GO) annotation in *D. melanogaster* orthologs, and performed a gene set enrichment test using for each gene family a score reflecting the overall occurrence of positive selection in the ant lineage (Materials and Methods; Supplementary text). Such an approach of grouping genes enables a more sensitive search for positive selection, while buffering the impact of potential false positives (e.g., from remaining alignment errors or GC-biased gene conversion events which are difficult to distinguish from real positive selection signals; see Supplementary text and Tables S20 to S24). Twenty-four functional categories of genes were significantly enriched for positively selected genes in the ant lineage (at 20% FDR; Table 3). A large number of them (11 out of 24) were related to mitochondria and mitochondrial activity. The other significant categories were related to nervous system development, behavior, immunity, protein translation and degradation, cell membrane, and receptor activity. Thus positive selection apparently targeted a diverse array of gene functions during the evolution of the ant lineage.

## *Usual targets of positive selection in insects*

To identify GO categories that experienced positive selection not only in ants but also in other insects, we reanalyzed the fly and bee datasets with the same methodology used for the ant dataset. These analyses revealed 106 GO categories significantly enriched for flies and 38 for bees (Tables 4 and 5; Tables S4 and S5). We investigated which categories were enriched for positively selected genes in the three lineages. The first group of genes commonly enriched in ants, flies and bees was related to proteolysis. This group included four of the 24 significantly enriched GO categories in ants ("proteolysis", "metallopeptidase activity", "peptidase regulator activity" and "hydro-lyase activity"), 8 of the 106 GO categories enriched in flies ("serine-type endopeptidase activity", "endopeptidase activity", "proteolysis", "metalloendopeptidase activity", "peptidase activity", "peptidase activity, acting on L-amino acid peptides", "metallopeptidase activity", "exopeptidase activity"), and 6 of the 38 GO categories enriched bees ("amine metabolic process", "metalloendopeptidase activity", "metallopeptidase activity", "signal transduction", "cellular amine metabolic process" and "cellular amino acid metabolic process").

The second group of genes enriched for positive selection signal in ants, flies and bees was involved in response to stimuli. There was an enrichment of the GO category "receptor activity" in the three lineages as well as the GO categories "transmembrane receptor activity" and "olfactory receptor activity" in flies. This class of genes plays a pivotal role in the interactions between individuals and their environment. In addition, the GO categories "response to biotic stimulus" and "response to other organism" were enriched in flies, and the GO category "response to hormone stimulus" was enriched in bees. In ants, "response to ecdysone" and "response to steroid hormone stimulus" were marginally significant (FDR = 21%; Table S6).

Some functions were enriched for positively selected genes in only two of the three lineages. These included GO categories related to immunity that were enriched in ants and flies, and some categories related to metabolism which were enriched in flies and bees. Evidence for positive selection on immunity-related functions in ants came from a significant enrichment of the GO categories "immune system development" and "hemopoietic or lymphoid organ development", the organ that produces during larval development the cells mediating the immune response in insects (Corley, Lavine 2006). Seven GO categories related to immunity were also enriched in flies ("defense response", "immune system process", "regulation of antimicrobial humoral response", "regulation of immune effector process", "regulation of defense response", "immune response" and "antimicrobial humoral response"). The absence of significant enrichment for related categories in bees might reflect a lack of power of the gene set enrichment test, because the set of immunity genes is small in the honeybee (Honeybee Genome Sequencing

Consortium. 2006) and the dataset analyzed was further depleted in genes with immunity-related functions (Supplementary text and Tables S7). Consistent with this interpretation, there was a trend in the direction of an enrichment, although non significant, for 5 of the 7 tested GO categories related to immunity in bees (data not shown), suggesting that immunity might be a common target of positive selection in insects.

The second set of GO categories enriched in two of the three insect lineages included various metabolic processes and their regulators, with metabolism of chitin, aminoglycan, carbohydrate, polysaccharide, glucose, hexose, glycerolipid, and phosphatidylinositol being enriched in flies and metabolism of lipid, amino-acid, nucleotide, and phosphorus being enriched in bees. There was no significant enrichment for GO categories related to metabolism in ants, but some categories were close to significance (e.g., "chitin metabolic process" and "rRNA metabolic process", with FDR=21% and 24% respectively; Table S6). Metabolic functions, such as amino-acid, fatty acid, lipid or RNA metabolism, were significantly enriched in ants when we used KEGG pathways annotation instead of the Gene Ontology to perform the gene set enrichment test, as well as when the single-copy orthologs dataset was reanalyzed with another multiple alignment method and a different quality filtering method (Table S25 and S26; Supplementary text). It thus seems that metabolism is a common target of positive selection in insects.

### *Social behaviors*

Two of the GO categories enriched in ants ("neuron recognition" and "adult locomotory behavior") might potentially be linked to the evolution of neural systems and behavior (Table 3). The first category was "neuron recognition". However, GO categories related to neural systems were also enriched in a non-social hymenoptera lineage ("regulation of synaptogenesis", "mushroom body development" and "memory" on branch #8, basal to the Hymenoptera; Figure 1, Table S9), and in the branches leading to primitively social lineages of bees ("synapse", "synapse organization", "regulation of synaptic growth at neuromuscular junction"; data not shown, (also reported in Woodard et al. 2011), suggesting that positive selection on neural system genes in ants might not be directly associated with the emergence of social behaviors.

The second enriched GO category enriched in ants was “adult locomotory behavior”. This category was not enriched in any of the other tested lineages. The three genes contributing most to the positive selection signal in this GO category were *DCX-EMAP, turtle* and *beethoven*. Mutational analyses of these genes in *Drosophila* suggest that they play an important role in sensory perception functions. Adult flies carrying a piggyBac insertion in *DCX-EMAP* are uncoordinated and deaf and display loss of mechanosensory transduction and amplification (Bechstedt et al. 2010). *Turtle* plays an essential role in the execution of coordinated motor output in complex behaviors in flies, notably regarding the response to tactile stimulation (Bodily et al. 2001). Finally, *beethoven* is involved in male courtship behavior, adult walking behavior and sensory perception of sound in flies (Tauber, Eberl 2001). This suggests that positive selection might have been important for the evolution of sensory perception functions in ants.

A specific analysis of olfactory receptor genes (ORs) did not provide support for the evolution of sociality being associated with increased levels of positive selection on ORs. A scan for positive selection across branches of a tree gathering 873 manually annotated ORs from two ants (*Pogonomyrmex barbatus* and *Linepithema humile*) and the solitary wasp *Nasonia vitripennis* (see Materials and Methods), revealed that positive selection was pervasive, with 277 branches (23%) displaying significant signals for positive selection (Figure 3, Figure S1). However, positive selection was detected in only 19% of the 929 branches leading to ant species, whereas as many as 40% of the 156 branches leading to wasps were under positive selection (Fisher exact test p=7.6e-9).

## *Phenotypic plasticity among castes*

None of the GO categories enriched for positively selected genes in ants could be linked to phenotypic plasticity (i.e., caste differences). In particular, there was no evidence of a significant enrichment for GO categories related to morphology or morphogenesis in the ant lineage. Another enrichment test using annotations obtained from mutant phenotypes in *D. melanogaster* (which are more relevant than GO annotations to analyze genes involved in morphogenesis since genes sets mostly refer to anatomical structures) also provided no clear support for positive selection on genes associated with phenotypic plasticity in ants (Table S12).

However, among the genes with the highest support for positive selection in ants (Table 1), two genes had a role in wing development (*Guanine nucleotide exchange* *factor in mesoderm* and *Methylthioribose-1-phosphate isomerase*) and one in larval development (*Dis3*), suggesting that even if positive selection did not act consistently on large sets of genes related to morphogenesis, it could have acted specifically on a few individual genes.

### *Mitochondrial genes*

Eleven GO categories enriched for positively selected genes in ants were related to mitochondrial activity (e.g., "mitochondrial electron transport", "mitochondrial matrix", "mitochondrial respiratory chain", "NADH dehydrogenase ubiquinone activity" and "oxidoreductase activity"; Table 3). The mitochondrial processes under positive selection were not restricted to respiration and energy production, but also included translation ("organellar ribosome", "mitochondrial small/large ribosomal subunit"). GO categories related to mitochondria were also enriched for positively selected genes on many individual branches of the ant lineage analyzed separately (Table S9), and in a larger dataset including duplicated genes analyzed with the site test (see Materials and Methods; Table 1, Table S13 and S19). This indicates that recurrent events of positive selection occurred on genes with mitochondrial functions during the evolution of the ant lineage. In contrast, mitochondria-related GO categories did not display any enrichment for positively selected genes in flies and bees (Tables 4 and 5), despite a high power to detect it on the respective datasets (Supplementary text). Similarly, no mitochondrial function was significantly enriched in the branches #7 and #8, basal to the ant lineage (Figure 1; Table S9), reinforcing the idea that increased positive selection on mitochondria is restricted to the ant lineage.

Of note, none of the 13 protein coding genes (Oliveira et al. 2008; Gotzek, Clarke, Shoemaker 2010) from the mitochondrial genome was included in our main dataset because the mitochondrial genomes of most of the ant species analyzed were not annotated. Our results thus reflect positive selection on nuclear genes encoding proteins which function in the mitochondrion. We annotated mitochondrial genomes in 5 of the 7 ant species analyzed and tested if positive selection could also be detected on the mitochondrial genomes themselves (Materials and Methods)(Gerber et al. 2001; Bazin, Glemin, Galtier 2006; Meiklejohn, Montooth, Rand 2007). However, we did not find evidence for positive selection on these alignments, neither with the branch-site test (Table S15) nor with the site test (Table S14).

### *Lifespan genes*

There was a significant enrichment for positively selected genes in the ant orthologs of *D. melanogaster* genes that were down-regulated in 61 days-old flies compared to 10 days-old flies, based on a published microarray analysis (p=0.011; below Bonferroni threshold $\alpha$=0.05/4=0.0125; Table S16)(Landis et al. 2004).

Two other genes known to be involved in aging were among the top-scoring genes for positive selection in our dataset. The first was *Tequila*, which has been shown to be associated with aging in an experimental evolution study in *D. melanogaster* (Remolina et al. 2012). The other was *Mitochondrial trifunctional protein α subunit,* whose knock-out also reduces lifespan in *D. melanogaster* (Kishita, Tsuda, Aigaki 2012). Although not in the list of top hits, *Sod2* (*Superoxide dismutase [Mn], mitochondrial),* a gene known to have antioxidant activity and whose overexpression has been shown to be associated with increased lifespan in some strains of *D. melanogaster* (Mockett et al. 1999; Curtis et al. 2007), underwent positive selection at the base of the Hymenoptera lineage (FDR=0.0073) and in the *Accromyrmex echinatior* branch (FDR=9.6e-8).

### *Selective pressure on genes with caste-biased expression*

There was a marginally significant enrichment for positively selected genes among genes with biased expression in adult workers in *S. invicta* (effect size=1.2, p=0.025; not significant after Bonferroni correction $\alpha$=0.05/6=0.0083; Table S17) and a stronger enrichment for genes with queen-biased expression in adults (effect size=1.8, p=0.0028). Surprisingly, however, there was a pattern of weaker enrichment for genes with male-biased expression in adults (effect size=1.04, p=0.2381). At the pupal stage, we did not detect a significant enrichment for positively selected genes among any group of genes showing caste-biased expression. But similarly to the adult stage, the enrichment effect size was higher for genes with queen-biased expression (effect size=1.2) than for genes with worker-biased expression (effect size=1.1), and it was the lowest for genes showing male-biased expression (effect size=1.06).

## Discussion

In this paper, we report results from a genome-wide scan for positive selection in protein coding sequences of seven ant genomes, using the rigorous branch-site model of Codeml (Zhang, Nielsen, Yang 2005) with stringent data quality control. Positive selection was detected in the ant lineage for 20% of the gene families analyzed. This proportion is similar in magnitude to the values observed in the other two insect lineages that we reanalyzed in this study: 20% in the 10 bee species, and 36% in the 12 flies species.

Our analysis identified similarities in patterns of positive selection between the ants and other insect lineages. Notably, at the broadest phylogenetic scale that our datasets allowed us to study, functional categories related to proteolysis, metabolism, response to stimuli, and immunity, were enriched for positively selected genes in ants, bees and flies. Interestingly, studies in mammals, fishes and urchins also provided evidence for positive selection on similar functional categories (Kosiol et al. 2008; Studer et al. 2008; Oliver et al. 2010; Montoya-Burgos 2011). Recurrent positive selection on such long evolutionary time scales is typical of genes involved in the interaction with changing environments or in conflict and competition, such as evolutionary arms races between sexes or between different species, which cause the perpetuation of adaptations and counter-adaptations in competing sets of co-evolving genes (Dawkins, Krebs 1979). It is notable that positive selection patterns on these categories of genes do not seem to reflect or be strongly affected by the large life-history differences between lineages analyzed here, for example the evolution of eusociality in the hymenopteran lineages. In particular, our results on immunity-related genes challenge the hypothesis that hygienic behaviors in social insects could have relaxed the selective pressure on immune genes, since this should be reflected in reduced levels of positive selection on these genes (Honeybee Genome Sequencing Consortium. 2006; Smith et al. 2008; Viljakainen et al. 2009; Werren et al. 2010; Fischman, Woodard, Robinson 2011; Smith et al. 2011a; Smith et al. 2011b; Suen et al. 2011; Harpur, Zayed 2013).

Our analysis indicated that genes involved in neurogenesis were under positive selection in ants and the primitively social lineages of bees. It was previously hypothesized that stronger selection on genes related to brain function and development should be observed in eusocial Hymenoptera species due to high cognitive demands associated with social life (Fischman, Woodard, Robinson 2011). However, our results are not consistent with this prediction because we also uncovered signs of positive selection at the base of the Hymenoptera lineage, that is before the evolution of sociality. Interestingly, a similar pattern had previously been reported with brain morphological data. A comparative analysis of insects showed that the size of mushroom body started to increase at the base of the Euhymenopteran (Orussioidea+Apocrita) lineage, approximately 90 Myr before the evolution of sociality in the Aculeata, and that there was no clear correlation between the size of brain components and the levels of sociality or cognition capabilities (Farris, Schulmeister 2011; Lihoreau, Latty, Chittka 2012). To account for this observation, Fischman and colleagues tried to identify factors, other than sociality, that may have placed unique selective pressure on brain evolution in species of the Hymenoptera lineage (Fischman, Woodard, Robinson 2011). Based on the observation that there was less positive selection on neurogenesis genes in highly social bees than in primitively social bees, they proposed that cognitive challenges might be associated with the mode of colony founding in social Hymenoptera. In particular, primitively social bees, which transit from a solitary phase during the process of colony founding, to a social phase could experience higher cognitive needs than highly social bees, which never go through a solitary phase. However, our results are also inconsistent with this model since increased positive selection was observed before the evolution of sociality in Hymenoptera. A comprehensive survey of positive selection on neurogenesis genes in Hymenoptera species, including species basal to the lineage, is required to identify precisely when the selective regime of these genes started to change, and in which hymenopteran sub-lineages it was maintained.

Our results also challenge the hypothesis that genes involved in chemical signaling experienced increased positive selection in social insects (Ingram, Oefner, Gordon 2005; Robertson, Wanner 2006; Bonasio et al. 2010; Smith et al. 2011a; Wurm et al. 2011; Zhou et al. 2012; Leboeuf, Benton, Keller 2013). The analysis of olfactory receptor repertoires in two ants and a non-social wasp indicate that positive selection on amino-acid substitutions was surprisingly less frequent in ant than in wasp branches. Given the limited number of species used in this analysis, future work should concentrate on generating extensive annotation of olfactory receptors from more Hymenoptera as well as outgroup species to identify characters or traits that could be associated with the pattern of positive selection on olfactory receptors.

Although our analyses did not provide support for previous hypotheses about the expected effect of social evolution on gene evolution, we identified several interesting functional categories which were enriched for positively selected genes exclusively in the ant lineage, possibly reflecting ant-specific adaptations. The most consistent and robust result was that genes functioning in the mitochondria were particularly likely to be under positive selection. Mitochondrial activity plays an important role in the process of reproductive isolation and speciation (Lee et al. 2008; Burton, Barreto 2012), interactions with endosymbionts such as *Wolbachia* (Werren 1997), diseases (Cortopassi 2002; Richly, Chinnery, Leister 2003; Trifunovic et al. 2004; Trifunovic et al. 2005), and in the process of aging (Lenaz 1998; Cortopassi 2002; Kowald, Kirkwood 2011). In that respect it is notable that the evolution of sociality has been accompanied by a nearly 100-fold increase in lifespan of queens compared to their solitary ancestors (Keller, Genoud 1997; Jemielity et al. 2005). Three lines of evidence suggest that increased lifespan of queens might be related to increased positive selection on mitochondrial genes in the ant lineage.

First, lifespan extension in insects, but also in other lineages such as birds and bats, appears to be associated with decreased production of Reactive Oxidative Species (ROS)(Perez-Campo et al. 1998; Brunet-Rossinni 2004; Parker et al. 2004; Corona et al. 2005; Jemielity et al. 2005). ROS are a normal by-product of cellular metabolism. In particular, one major contributor to oxidative damage is hydrogen peroxide ($H_2O_2$), which is produced from leaks of the respiratory chain in the mitochondria (Harman 1972; Lenaz 1998; Finkel, Holbrook 2000; Cui, Kong, Zhang 2012). Positive selection in ants on genes functioning in the mitochondria may thus reflect selection to increase mitochondrial efficiency and reduce ROS production. Interestingly, positive selection on genes with mitochondrial functions was previously documented in the bat lineage (Shen et al. 2010; Zhang et al. 2013), which include species with exceptional longevity (Brunet-Rossinni, Austad 2004). In the bat *Myotis lucifugus,* ROS production was also shown to be significantly lower than in two similar sized mammal species (a mouse and a shrew) although the metabolic rates, and thus mitochondrial activity, of the former were much higher because of flight demands (Brunet-Rossinni 2004).

Second, on the basis of gene expression data obtained in the fire ant *S. invicta*, our analyses revealed that positive selection was strongest on genes with queen-biased expression, intermediate on genes with worker-biased expression, and weakest on genes with male-biased expression. This association between levels of positive selection and caste-biased differences in gene expression cannot be simply accounted by differences in expression levels of mitochondrial genes (which are enriched for positively selected genes in ants) since in *S. invicta* mitochondrial genes are significantly less expressed in queens than in workers at the larval stage, and not differentially expressed at the adult stage (Figure S2). The finding of higher levels of positive selection for genes more highly expressed in the castes with the longer lifespan (queens can live decades in some species, whereas workers have lifespan in the order of months, and males in the order of days) suggests that increased positive selection on queen-specific genes could be related to longer lifespan.

Third, our analyses showed that the levels of positive selection were higher on orthologs of genes which are down-regulated during aging in flies. These genes include numerous energy metabolism genes, and their down-regulation in old flies is thought to reflect a decline of normal and functional mitochondria with age (Yui, Ohno, Matsuura 2003; Landis et al. 2004). The finding of increased levels of positive selection on genes whose expression declines at older ages suggests that the function of these genes might be improved in ants, potentially delaying the loss of normal activity in mitochondria with age. It would be interesting to test if parallel mechanisms also evolved in the ant lineage to maintain the expression of these genes and delay the decline of mitochondria activity through lifespan in queens.

In contrast to ants, there was no evidence of elevated levels of positive selection on mitochondrial functions in bees. As most social species, bees also evolved longer queen lifespans (more than 2 years) compared to males and workers (a few weeks)(Keller, Genoud 1997; Munch, Amdam, Wolschin 2008). There are four possible explanations for the difference between ants and bees in the level of positive selection on mitochondrial genes. First, lifespan differences between castes are less pronounced in bees, where queens live up to 2-5 years, than in ants, where queens can live up to 30 years, possibly resulting in lower selective pressure to increase lifespan in bees than in ants. Second, because eusociality evolved independently in ants and bees it is possible that extended queen lifespans evolved by different molecular mechanisms (Jemielity et al. 2005; Jobson, Nabholz, Galtier 2010). For example, vitellogenin may play a more central role for aging in bees than ants (Amdam, Omholt 2002; Corona et al. 2007; Munch, Amdam, Wolschin 2008). Third, the evolution of mitochondria-related genes may have been differently constrained in ants and bees. For example, metabolic rates differ greatly between flying bee workers and non-flying ant workers because flight is an energetically costly behavior requiring highly elevated metabolic rates (Jensen, Holm-Jensen 1980; Suarez 2000; Niven, Scharlemann 2005). Because metabolism and mitochondrial activity are closely connected, lower metabolic rates in ants might have alleviated functional constraints on mitochondria-related genes, allowing selection to act on lifespan extension. Fourth, the GC content in bee genomes was shown to be lower than in ant genomes (Honeybee Genome Sequencing Consortium. 2006; Simola et al. 2013). Some parts of the bee genomes, in particular their mitochondrial genomes (Crozier, Crozier 1993; Gotzek, Clarke, Shoemaker 2010; Tan et al. 2011), display extreme bias in nucleotide composition, which leads to significant effect on both the codon usage patterns and amino acid composition of proteins and may have interfered with the action of positive selection.

If positive selection acted to optimize the functioning of mitochondria in ants, it could be expected that the mitochondrial genome itself should be targeted by positive selection. However, mitochondrial genes generally exhibit very low $d_N/d_S$ ratios (Montooth, Rand 2008) and there was no clear evidence in our results for positive selection on the 13 genes of the mitochondrial genome itself. This suggests that innovations related to mitochondrial activity could arise more easily on nuclear genes, whereas mitochondrial genes seem more likely to maintain conserved core functionalities.

In conclusion, this study provides a detailed analysis of the extent of positive selection events on protein-coding genes in seven ant species. Because false positives are a major concern for whole-genome scans for positive selection, we used a conservative methodology. We also reanalyzed data in bees and flies with the same methods to permit an unbiased and robust comparison of positive selection between lineages. The comparison between these three lineages provided interesting perspectives on the evolution of genes implicated in immunity, neurogenesis and olfaction, and allowed us to pinpoint positive selection events that were specific to the ant lineage. In particular, we found that the evolution of extreme lifespan in ants was associated with positive selection on genes with mitochondrial functions, suggesting that a more efficient functioning of mitochondrial genes might have been an important step towards the extreme lifespan extension that characterizes this lineage. It would be interesting to complement this study by scans for genes under lineage-specific strong or relaxed purifying selection, to get a more global picture of natural selection patterns in ant genomes, and uncover additional genes that could have played a significant role during the evolution of the ant lineage.

# Materials and Methods

## *Single-copy orthologs gene families dataset*

Protein coding gene sequences of the 7 ant genomes were downloaded from the Hymenoptera Genome Database (http://hymenopteragenome.org/ant_genomes/)(Munoz-Torres et al. 2011).

The complete annotated gene sets were OGS_1.0 for *Acromyrmex echinatior* (Nygaard et al. 2011), OGS_1.2 for *Atta cephalotes* (Suen et al. 2011), OGS_2.2.3 for *Solenopsis invicta* (Wurm et al. 2011), OGS_1.2 for *Pogonomyrmex barbatus* (Smith et al. 2011b), OGS_3.3 for *Camponotus floridanus* (Bonasio et al. 2010), OSG_1.2 for *Linepithema humile* (Smith et al. 2011a) and OGS_3.3 for *Harpegnathos saltator* (Bonasio et al. 2010). Coding sequences of 5 outgroup species were downloaded from the Hymenoptera Genome Database for the honey bee (*Apis mellifera* Amel_pre_release2)(Honeybee Genome Sequencing Consortium. 2006) and the jewel wasp (*Nasonia Vitripenis* OGS_v1.2)(Werren et al. 2010), from Flybase (Tweedie et al. 2009) for the fruit fly (*Drosophila melanogaster* FB5.29)(Adams et al. 2000), from BeetleBase (Kim et al. 2010b) for the flour beetle (*Tribolium castaneum* Tcas_3.0)(*Tribolium* Genome Sequencing Consortium. 2008), and from vectorBase (Lawson et al. 2009) for the body louse (*Pediculus humanus* PhumU1.2)(Kirkness et al. 2010).

Gene families were obtained from a custom run of the OrthoDB pipeline for the Ant Genomic Consortium (http://cegg.unige.ch/orthodbants and http://bioinfo.unil.ch/supdata/Roux_positive_selection_ants/orthoDB_run.zip; pipeline of OrthoDB release 4)(Waterhouse et al. 2011; Simola et al. 2013). Briefly, OrthoDB implements a Best Reciprocal Hit clustering algorithm based on all-against-all Smith-Waterman protein sequence comparisons. The longest alternatively spliced form of genes is used. The orthologous groups are built at different taxonomic levels and it is possible to query for specific phyletic profiles by combining the criteria of absent, present, single-copy, multi-copy or no restriction, for each species within the studied clade.

Gene families including strictly one ortholog in each of the 12 species were selected (2,756 gene families). Because annotations of the studied genomes are likely to be incomplete (Simola et al. 2013), families with a few missing genes – gene losses or unannotated genes – were included, with the restriction that at least 4 genes out of the 7 ant species, and 3 genes out of the 5 outgroup species should be present in the gene family. Simola et al. (2013) have shown that among the seven ant species, there were generally few lost or missing genes, apart from *S. invicta* (less than 400 *S. invicta* genes were missing in single-copy orthologs families) and *A. echinatior* (less than 150 *A. echinatior* genes were missing in single-copy orthologs families). Our gene family selection criteria allow handling such a moderate amount of missing genes in families. In order to transfer functional annotations from *D. melanogaster*, only families including a fruit fly ortholog were retained. With these criteria, the number of OrthoDB groups in the dataset increased to 4,337. All gene families were assumed to follow the species tree topology (Figure 1). The exclusion of families that experienced gene duplication facilitates the comparison of branches between gene families, and keeps our analysis from biases related to differential duplication among lineages (Waterhouse, Zdobnov, Kriventseva 2011) and among genes (Davis, Petrov 2004; He, Zhang 2006), and to the consequences of duplication (Force et al. 1999; Brunet et al. 2006). Finally, results on single-copy orthologs can be easily compared to previously published studies using similar gene family topologies (Drosophila 12 Genomes Consortium 2007; Kosiol et al. 2008; Studer et al. 2008; Lindblad-Toh et al. 2011).

Basic sequence quality features were first controlled as in (Hambuch, Parsch 2005). CDS (coding sequences) whose length was not a multiple of three or did not correspond to the length of the predicted protein, or that contained an internal stop codon, were eliminated; the longest CDS of genes showing multiple isoforms was retained; CDS shorter than 100 nt were eliminated.

Because misalignment errors can be an important source of false positives in genome-wide scans for positive selection in coding sequences (Schneider et al. 2009; Markova-Raina, Petrov 2011; Yang, dos Reis 2011; Jordan, Goldman 2012), we took great care at filtering the potentially problematic sites in the alignments. The quality filtering pipeline used here is adapted from the pipeline of the Selectome database release 4 (http://selectome.unil.ch)(Proux et al. 2009; Moretti et al. 2014). The multiple alignment of the protein sequences in each gene family was computed by M-Coffee (Wallace et al. 2006) from the T-Coffee package v8.93 (Notredame, Higgins, Heringa 2000), which combines the output of different aligners. Similarly to Ensembl Compara (see http://www.ensembl.org/info/docs/compara/homology_method.html for more

details)(Vilella et al. 2009), 4 different aligners were used for M-Coffee (mafftgins_msa, muscle_msa, kalign_msa and t_coffee_msa). M-Coffee outputs a consensus of 4 alignments from the different aligners, and a quality score for each residue based on the concordance of the alignment at each position by different aligners. Scores lie between 0, if a residue was not aligned at the same position by the different aligners, and 9 if it is reliably aligned at the same position in all cases. Reliably aligned residues with a score of 7 or above were retained. We used the heuristic algorithm of MaxAlign v1.1 (Gouveia-Oliveira, Sackett, Pedersen 2007) to detect and remove sequences badly aligned as a whole (gap-rich sequences) in the multiple sequence alignments. When a sequence was removed, the gene family was realigned and refiltered using M-Coffee. Families left with less than four sequences were discarded because of insufficient power to detect positive selection. The protein alignments were reverse-translated to nucleotide alignments using the seq_reformat utility of the T-Coffee package (Notredame, Higgins, Heringa 2000).

We used a stringent Gblocks filtering (v0.91b; type = codons; minimum length of a block = 4; no gaps allowed)(Talavera, Castresana 2007) to remove gap-rich regions from the alignments, as these are problematic for positive selection inference (Fletcher, Yang 2010; Markova-Raina, Petrov 2011). The large memory requirements of M-Coffee for long alignments led us to use only Gblocks without M-Coffee scoring if the length of the alignment was greater than 9,000 nt.

After filtering, our dataset included 4,261 gene families with an average of 10.4 branches per family to test (Figure 1; 44,306 branches to test; median = 10 branches per family). The mean length of filtered alignment was 1,133 nt. (median = 885 nt.), ranging from a minimum of 54 nt. to a maximum of 22,248 nt. Of note, lost or missing genes in families affect the topology of the trees and the possibility to compare equivalent branches of different families. In total, our dataset contains 36,681 branches (83%) in 4,256 families which corresponded to the canonical topology defined by the species tree (Figure 1) and could be compared across families (e.g., Table 2).

Our analyses are likely to underestimate the genome-wide number of positive selection events because i) single-copy orthologs tend to evolve under stronger purifying selection than multi-copy gene families (Waterhouse, Zdobnov, Kriventseva 2011), ii) the ant genomes still lack good annotation of gene models and single-copy orthologs gene families could be missed, iii) we filtered out unreliable parts of sequence alignments including fast evolving residues that are difficult to align (Fletcher, Yang 2010; Privman, Penn, Pupko 2012). The last point is balanced by the fact that conserved regions might be more prone to positively selected substitutions (Bazykin, Kondrashov 2012) and that the removal of unreliable regions seems to increase the power to detect positive selection (Jordan, Goldman 2012; Privman, Penn, Pupko 2012).

### *Extensive gene families dataset*

Another dataset gathered all gene families from the OrthoDB database that could pass our quality filters, and notably families that experienced duplications. The CDS were filtered as described above. Amino-acid sequences were aligned using PAGAN version 0.47 (Loytynoja, Vilella, Goldman 2012). The program GUIDANCE (v1.1) was used to assess alignment confidence and mask unreliably aligned residues (Penn et al. 2010; Privman, Penn, Pupko 2012). The combination of a phylogeny-aware aligner (PAGAN replaces PRANK (Löytynoja, Goldman 2008) and is based on the same principle) and of this filtering algorithm was shown to perform the best in recent benchmark studies on simulated data (Jordan, Goldman 2012; Privman, Penn, Pupko 2012). Gene family phylogenies were built using RAxML (v7.2.9)(Stamatakis 2006) from the amino acid sequences, with the LG matrix and the CAT model. Amino acid alignments were reverse-translated into the corresponding codon alignments. This resulted in 6,186 families tested, with an average of 11 genes, and an average length of filtered alignment of 3,129 nt. (median of 2,385 nt., ranging from a minimum of 192 nt. to a maximum of 20,556 nt.).

### *Mitochondrial gene families dataset*

Contigs corresponding to mitochondrial genomes could be downloaded for 5 of the 7 ant genomes (*Acromyrmex echinatior, Atta cephalotes, Solenopsis invicta, Pogonomyrmex barbatus* and *Linepithema humile*). They were submitted to MITOS, a web server for the annotation of metazoan mitochondrial genomes (http://mitos.bioinf.uni-leipzig.de/index.py)(Bernt et al. 2012). This gave us the predicted coordinates of 13 mitochondrial protein coding genes in these species. Frameshift errors or incomplete gene predictions were manually corrected. Mitochondrial genes from the outgroup species *Apis mellifera*, *Nasonia Vitripenis* and *Tribolium* castaneum were downloaded from Genbank (accessions L06178; EU746609.1 and EU746613.1; AJ312413.2 and NC_003081.2 respectively). Mitochondrial genes from *Drosophila melanogaster* were downloaded from Flybase at ftp://ftp.flybase.net/genomes/Drosophila_melanogaster/dmel_r5.43_FB2012_01/fasta/dmel-dmel_mitochondrion_genome-CDS-r5.43.fasta.gz. The alignment and filtering steps for the 13 mitochondrial gene families were identical to the dataset of single-copy orthologs nuclear gene families (see above). A total of 119 branches were tested in this dataset (average of 9.2 and median of 9 branches per family; average length of filtered alignment of 641 nt. and median of = 621 nt., ranging from a minimum of 39 nt. to a maximum of 1,413 nt.).

## *Twelve Drosophila dataset*

Single-copy ortholog gene families from the 12 sequenced Drosophila species were downloaded from ftp://ftp.flybase.net/12_species_analysis/clark_eisen/alignments/ (files "all_species.guide_tree.longest.cds.tar.gz" and "all_species.guide_tree.longest.translation.tar.gz")(Drosophila 12 Genomes Consortium 2007). The alignment and filtering steps for these gene families were identical to the dataset of single-copy ortholog gene families used for the ant analysis. Out of 6,698 initially downloaded Drosophila gene families, 3,749 (56%) passed our filters and could be tested for positive selection, resulting in 77,495 branches tested (average of 20.7 and median of 21 branches per family; average length of filtered alignment of 876 nt. and median of 708 nt., ranging from a minimum of 15 nt. to a maximum of 14,535 nt.).

## *Bee dataset*

Single-copy ortholog gene families from 10 bee species were downloaded from http://insectsociogenomics.illinois.edu/. This set of gene families is incomplete as it is derived from the sequencing of expressed sequence tags (using 454 Life Science/Roche GS-FLX platform) from 9 bee species (Woodard et al. 2011), and from gene models of the honey bee *Apis mellifera* (Honeybee Genome Sequencing Consortium. 2006). The alignment and filtering steps for these gene families were identical to the dataset of single-copy ortholog gene families used for the ant analysis. Out of 3,647 initially downloaded gene families, 2,256 (62%) passed our filters and could be tested for positive selection, resulting in 20,169 branches tested (average of 8.9 and median of 9 branches per family; average length of filtered alignment of 611 nt. and median of 528 nt., ranging from a minimum of 27 nt. to a maximum of 3,945 nt.).

## *Branch-site test for positive selection*

We used the updated branch-site test (Zhang, Nielsen, Yang 2005) of Codeml from the package PAML v4.4c (Yang 2007) to detect Darwinian positive selection experienced by a gene family in a subset of sites in a specific branch of its phylogenetic tree. This test was previously used in genome-wide scans for positive selection in various lineages (Bakewell, Shi, Zhang 2007; Kosiol et al. 2008; Studer et al. 2008; Vamathevan et al. 2008; Oliver et al. 2010; George et al. 2011) and is used by the Selectome project (http://selectome.unil.ch)(Proux et al. 2009; Moretti et al. 2014). It is acknowledged to be more sensitive for the detection of positive selection than branch tests (Yang 1998) or site tests (Yang et al. 2000), because it does not average the signal over all codons in the alignment (branch test) nor over all branches of the phylogeny (site test)(Yang, dos Reis 2011). It is also robust to relaxation of purifying selection ($\omega$ close to 1) since this scenario is accounted for in the null model (Zhang 2004; Zhang, Nielsen, Yang 2005). The alternative model is contrasted to the null model using a likelihood-ratio test (LRT), where log-likelihood ratios are compared to a chi-square distribution with 1 degree of freedom (Zhang, Nielsen, Yang 2005). Previous studies have reported the branch-site test to be conservative in this setup (Bakewell, Shi, Zhang 2007; Studer et al. 2008; Fletcher, Yang 2010; Yang, dos Reis 2011; Gharib, Robinson-Rechavi 2013). We did not use the $\omega$ estimates to infer the strength of positive selection because they were shown to be unreliable (Bakewell, Shi, Zhang 2007; Yang, dos Reis 2011).

In the absence of a specific *a priori* hypothesis regarding which branches to test for positive selection, our implementation runs the test multiple times on each gene family, successively changing the branch selected as foreground. The branches considered as foreground are highlighted in red in Figure 1. This approach was shown to be legitimate if p-values from the successive tests are corrected for multiple testing (Anisimova, Yang 2007; Yang, dos Reis 2011). We applied a False Discovery Rate (FDR) correction (Benjamini, Hochberg 1995) over all the p-values treated as one series (number of branches tested × number of gene families tested). In the ant single-copy orthologs nuclear dataset we analyzed a maximum of 15 branches leading to the 7 ant species, summing to 44,306 tests performed. In the ant mitochondrial dataset we analyzed a maximum of 11 branches leading to 5 ant species, summing to 119 tests (branches in red in Figure S3). In the Drosophila single-copy orthologs dataset we analyzed a maximum of 21 branches, leading to a total of 77,495 tests (Figure S4).

Finally in the bee dataset we analyzed a maximum of 17 branches, leading to a total of 20,169 tests (Figure S5).

All computations were performed using Slimcodeml (release 4th May 2011)(Schabauer et al. 2012), an optimized version of Codeml, based on the release 4.4c of the PAML package (downloadable at http://selectome.unil.ch/cgi-bin/download.cgi). Slimcodeml was estimated to run the branch-site models about 1.77 times faster than the original Codeml thanks to the use of external standard libraries for linear algebra calculations and specific optimizations for the computer architecture used. We verified on a subset of the gene families that the results given by Slimcodeml were identical with the original Codeml. Examples of Slimcodeml/Codeml control files used are provided in Supplementary text. For the ant mitochondrial dataset, Codeml was used with the option "icode=4" to use the Invertebrate mitochondrial genetic code (http://www.ncbi.nlm.nih.gov/Taxonomy/Utils/wprintgc.cgi#SG5).

The branch-site model is known to display convergence problems in the calculation of likelihoods (Yang, dos Reis 2011), leading to negative or artificially large log-likelihood ratios. We thus launched 3 independent runs for both the alternative and null hypotheses, for each branch of each gene family, and kept the best likelihood value of each run to calculate the log-likelihood ratio (Yang, dos Reis 2011). Of note, the likelihood differences observed across the 3 runs were most of the time very small. Even after reconciliation of 3 replicate runs, we still observed a number of negative log-likelihood ratios (8% of the tests – most of them very close to 0). In such cases we manually set the log-likelihood ratios to 0 (meaning non-significance). We recorded the largest differences in likelihood values between the three independent runs in both fixed and alternative models ($d$). The distribution of differences was bimodal, with a first major mode at $d$=0, gathering most data, and a second minor mode at $d$~1. A cutoff at $d$=0.004 clearly separated the 2 peaks. We used this stringent cutoff ($d$>0.004) to eliminate all tests with potential convergence problems in the fixed and alternative models (see Supplementary text and Table S23).

Values of $d_N$ and $d_S$ were calculated with parameters extracted from Codeml results files (.mlc files).

All calculations were performed on the SIB Vital-IT cluster in Lausanne (http://www.vital-it.ch/). All 3 runs and the 2 hypotheses of each test were performed on the same node of the cluster.

### *Site test for positive selection*

The site test (Yang et al. 2000) of Codeml from the package PAML v4.4e (Yang 2007), allowing the $d_N/d_S$ ratio ($\omega$) to vary among sites, was run on the extensive dataset of 6,186 families (see above). We contrasted the null model M8a (beta & $\omega$ with $\omega$=1) to the alternative model M8 (beta & $\omega$ with $\omega \geq 1$) with 11 site classes (Swanson, Nielsen, Yang 2003; Wong et al. 2004). Examples of Codeml control files used are provided in Supplementary text. Similar to the branch-site test, we launched 3 independent runs for both the alternative and null hypotheses for each gene family and kept the best likelihood value of each run for the likelihood ratio test (Table S19). The likelihood ratios were compared to a Chi-square distribution with 1 degree of freedom as recommended in PAML user's guide (http://abacus.gene.ucl.ac.uk/software/pamlDOC.pdf).

### *Reconstruction of ancestral G+C content*

The program nhPhyml (Galtier, Gouy 1998; Guindon, Gascuel 2003; Boussau, Gouy 2006) was used to estimate the G+C content at third codon positions at each node of the gene family trees (topology fixed, transition/transversion ratio estimated, alpha parameter estimated with eight categories). Following Studer et al. (2008), we calculated the shift in GC3 content at each branch as the difference between GC3 contents at the nodes delimitating that branch.

### *Olfactory receptors family*

Olfactory receptors are difficult to process in automated pipelines since they are characterized by dynamic patterns of duplications and pseudogenization during evolution (Nozawa, Nei 2007). Furthermore, the sequences of ORs are highly variable and notoriously difficult for automatic gene annotation. Accordingly, our main dataset of single-copy orthologs was depleted in genes involved in olfaction (Tables S7, S8, S10 and S11) and GO categories related to olfaction could not be tested for enrichment of positively selected genes because they included too few annotated genes. We therefore used a more comprehensive dataset of 873 manually annotated protein coding sequences of OR genes (excluding suspected pseudogenes) provided by Hugh Robertson for *P. barbatus* (291 genes)(Smith et al. 2011b), *L. humile* (320 genes)(Smith et al. 2011a) and *N. vitripennis* (262 genes)(Werren et al. 2010). Nucleotide sequences were translated and amino acid sequences were aligned using MAFFT (Katoh et al. 2005). Unreliably aligned residues were masked using GUIDANCE based on 32 bootstrap samples and a cutoff of 0.2 that was chosen so that the 15% lowest scoring residues are masked (Penn et al. 2010; Privman, Penn, Pupko 2012). Phylogeny was reconstructed using RAxML with the JTT substitution matrix, the CAT approximation and 100 bootstrap samples (Stamatakis 2006). Because the resulting gene tree was too large for an analysis with the branch-site test of Codeml, we divided it into 16 smaller subtrees, each containing less than 100 leaves. Branches with as high as possible bootstrap support were chosen as splitting points. The 16 subtrees include all ant sequences but only 105 *N. vitripennis* sequences. The sequences from each subtree were realigned using PRANK version 100701 (Löytynoja, Goldman 2008) and reverse-translated into corresponding codon alignments. GUIDANCE was used to mask unreliably aligned codons (0.8 cutoff). Phylogeny was reconstructed using RAxML as above. Out of 1,744 branches in the initial tree, 1,400 branches from the subtrees were tested using the branch-site test of Codeml (see above), and the computation was successful (both null and alternative hypotheses) for 1,184 branches. Significant branches are highlighted in red in Figure 3 and in Figure S1. Full results of the branch-site test on all 16 clades are shown in Table S18. A full tree with branch names and bootstrap values is provided as figure S1. Newick trees of the 16 individual subtrees along with annotation of tested branches are available in Supplementary text.

### *Tests of functional category enrichment*

Gene Ontology (GO)(Ashburner et al. 2000) annotations for gene families were taken from the annotation of the *D. melanogaster* gene member they include (downloaded from http://flybase.org/static_pages/downloads/FB2011_02/go/gene_association.fb.gz). The annotation of children GO categories was propagated to their parent categories following the GO graph structure. Gene Ontology categories mapped to 10 genes or less were discarded for the enrichment analysis.

To identify over and under-represented functional categories present in the datasets used in this study, the package topGO version 2.4 (Alexa, Rahnenfuhrer, Lengauer 2006) of Bioconductor (Gentleman et al. 2004) was used. A Fisher exact test was used, combined with the "elim" algorithm of topGO, which decorrelates the graph structure of the Gene Ontology to reduce non-independence problems (Alexa, Rahnenfuhrer, Lengauer 2006). The reference set was constituted of all OrthoDB families including a *D. melanogaster* gene with GO annotation. Gene ontology categories with a FDR < 20% are reported (Benjamini, Hochberg 1995).

Regarding the functional enrichment of genes targeted by positive selection, the Fisher exact test approach has been criticized because it imposes the choice of an arbitrary cutoff to dichotomize genes into "significant" and "non-significant" categories. This leads to a loss of information and limits the power and robustness of this method (Allison et al. 2006; Tintle et al. 2009; Daub et al. 2013). To test for GO functional categories for enrichment for positively selected genes, we instead used a gene set enrichment approach, which tests whether the distribution of scores of genes from a gene set differs from the whole dataset scores distribution, allowing the detection of gene sets that contain many marginally significant genes. Different implementations for this approach have been proposed. The most widely used is the Gene Set Enrichment Analysis (GSEA)(Subramanian et al. 2005), but it was shown to perform relatively poorly (Kim, Volsky 2005; Efron, Tibshirani 2007; Tintle et al. 2009). Here, we used a SUMSTAT test: for a given gene set *g* including *n* genes, the SUMSTAT statistic is defined as the sum of scores of the *n* genes. This statistic was shown to be more sensitive than a panel of other methods, while controlling well for the rate of false positives (Ackermann, Strimmer 2009; Tintle et al. 2009; Fehringer et al. 2012; Daub et al. 2013). To be able to use the distribution of log-likelihood ratios of the positive selection test – which follows a chi-square distribution with 1 degree of freedom and spans several orders of magnitude – as scores in the SUMSTAT test, we applied a fourth root transformation as variance stabilizing method. This transformation conserves the ranks of gene families (see http://udel.edu/~mcdonald/stattransform.html)(Canal 2005; McDonald 2009). According to the Central Limit Theorem the distribution of SUMSTAT scores from random gene sets approaches a normal distribution whose mean and variance derives from the mean and variance of the scores of the complete list of tested genes *G*:

$$E(SUMSTAT) = n \cdot E(G)$$

and

$$Var(SUMSTAT) = n \cdot Var(G)$$

We performed bidirectional tests against this distribution to test if the SUMSTAT statistic for a given gene set is higher or lower than expected by chance, corresponding to respectively enrichment or depletion for positively selected genes in this gene set. We verified the accuracy of this methodology by drawing an empirical null distribution for each gene set of size *n* found in the real dataset, based on scores of 10,000 gene sets of same size *n* randomly picked from the whole dataset. The distribution of SUMSTAT scores of these randomized gene sets approximates closely a normal distribution, even when the set size is small (Figure S6). This makes the SUMSTAT test less computationally intensive than other gene set enrichment approaches (e.g. GSEA)(Subramanian et al. 2005) where the null distribution cannot be inferred mathematically and randomizations have to be performed for each individual test. We verified that a GSEA approach gave broadly similar results (not shown).

Because different gene sets sometimes share many genes in common, the list of significant gene sets resulting from enrichment tests is usually highly redundant. We implemented the “elim” algorithm from the Bioconductor package topGO, to decorrelate the graph structure of the Gene Ontology (Alexa, Rahnenfuhrer, Lengauer 2006). Briefly, the GO categories are tested recursively starting from the deeper levels of the GO tree, and the genes annotated to these significant categories are removed from all their parent categories. As the tests for different categories are not independent, it is not clear if classical approaches to assess the False Discovery Rate (e.g., Benjamini, Hochberg 1995) are accurate. Thus we calculated empirically an FDR at each p-value threshold by performing 100 randomizations where the scores of gene families were permuted and the gene set enrichment test rerun. The FDR is estimated as:

$$FDR = \frac{FP}{FP + TP} = \frac{N_0}{N_t}$$

where at a given p-value threshold $N_0$ represents the mean number of false positives obtained in the randomizations and $N_t$ represents the number of positives obtained with the real dataset. The FDR obtained with this approach was in good agreement with the Benjamini-Hochberg FDR (Benjamini, Hochberg 1995). Gene Ontology categories with a FDR < 20% are reported. Functional categories depleted in positive selection reflect the most conserved sets of functional categories, under the action of purifying selection. These are not discussed in this manuscript.

The gene set enrichment test ran on each individual branch of the tree with results of the branch-site test yields heterogeneous results, probably resulting from differences in power of the branch-site test on different branches of the phylogeny (Table S9; only branches Sinv, Pbar, Hsal, #3 and #6 show some significant categories at FDR 20%). This test could also be sensitive to false positive results of the branch-site test (e.g., GC-biased gene conversion, discussed in Supplementary text). Thus we designed a test less sensitive to these problems. We considered a unique score per gene family reflecting the evidence for positive selection globally in the ant lineage, the mean of the branch-site test scores on the 13 individual ant branches. This scoring scheme should unveil functional categories of genes that experienced extensive and probably recurrent episodes of positive selection in the ant lineage, but is not strictly equivalent to using the results of a site test on ants branches, since it allows the detection of gene families with positive selection events affecting different sites on different branches. We also checked that in most cases, the enriched categories were not significant only because of a single outlier gene with a strong positive selection score, but displayed a significant shift in the distribution of positive selection scores of numerous genes (Figure S7).

Finally, as a sanity check, the gene set enrichment test was also performed using KEGG pathways annotation. KEGG pathways and the mapping to *D. melanogaster* genes were downloaded with the KEGG REST API (http://www.kegg.jp/kegg/rest/keggapi.html). Because hierarchical relationships among KEGG pathways are limited, we did not use the "elim" decorrelation algorithm. Pathways mapped to more than 10 genes were retained. In total 51 KEGG pathways were tested.

## *Tests of phenotypic category enrichment*

Mutant phenotype annotations of *D. melanogaster* genes were extracted from Flybase (Drysdale 2001; Osumi-Sutherland et al. 2013). The following ontologies were downloaded from the OBO foundry (Smith et al. 2007): the Flybase controlled vocabulary ontology (http://www.berkeleybop.org/ontologies/obo-all/flybase_vocab/flybase_vocab.obo), the Drosophila anatomical ontology (http://www.berkeleybop.org/ontologies/obo-all/fly_anatomy/fly_anatomy.obo) and the Drosophila developmental stages ontology (http://www.berkeleybop.org/ontologies/obo-all/fly_development/fly_development.obo). The relationships between genes and alleles, and between alleles and phenotypes (anatomical and developmental ontology categories) were extracted from Flybase (ftp://ftp.flybase.net/releases/FB2012_01/reporting-xml/FBgn.xml.gz; "derived_pheno_class" and "derived_pheno_manifest" entities). The information on gain or loss-of-function alleles was extracted from the file ftp://ftp.flybase.net/releases/FB2012_01/reporting-xml/FBal.xml.gz (loss of function: controlled vocabulary term FBcv:0000287 and child terms; gain of function: FBcv:0000290 and child terms). The annotation of child phenotypic categories (anatomy of development) was propagated to their parent categories following the respective ontologies structures.

To perform an enrichment analysis based on mutant phenotypes in fruit fly, we used the SUMSTAT test. Because the annotation is scarcer than the Gene Ontology annotation, we used only the categories mapped to more than 5 genes for the enrichment analysis. The reported results include the annotation for gain and loss-of-function alleles. We observed very similar results when gain-of-function alleles were removed from the annotation (Weng, Liao 2011)(not shown).

## *Expression data*

Microarray expression data from *Solenopsis invicta* (Ometto et al. 2011) were provided by the authors upon request. These included expression levels of clones of the spotted microarray used, as well as the list of genes identified to be over-expressed in each of the three castes (workers, queens and males), both at pupal and adult stages. The mapping of clones to the gene model of *Solenopsis invicta* (OGS_2.2.3)(Wurm et al. 2011) was provided by Y. Wurm, and is similar to the mapping used in (Hunt et al. 2011). If multiple clones mapped to the same gene, the average signal was used for expression. For differential expression, we used the results of the original study (BAGEL analysis, where a clone was considered to be differentially expressed between conditions if the Bayesian posterior probability was $p<0.001$, corresponding to a FDR~5%)(Ometto et al. 2011). A gene was considered differentially expressed if at least one clone mapped to it was found differentially expressed. Expression data was available for 1,327 genes of the single-copy orthologs dataset, including 603 genes over-expressed in at least one condition. We ran a SUMSTAT gene set enrichment test on the sets of genes with caste-specific expression (pupal male, pupal queen, pupal worker, adult male, adult queen and adult worker). P-values were obtained by comparison to an empirical distribution created with 10,000 randomizations of gene scores.

## *Aging genes*

Aging and oxidative stress associated genes were obtained from a microarray study in *Drosophila melanogaster* comparing the expression of genes in 10 day old flies to 61 day old flies, and flies exposed to 100% $O_2$ for 7 days to controls (Landis et al. 2004). We tested the enrichment for positively selected genes (SUMSTAT test) in four gene sets to constituted of up and down-regulated genes in both contrasts. P-values were obtained by comparison to an empirical distribution created with 10,000 randomizations of gene scores.

## *Genes with mitochondrial function*

Genes with mitochondrial function were identified as those mapped to any of the 310 Gene Ontology categories including "mitochondria*" in their names or synonym names (using the search engine on http://amigo.geneontology.org/). 313 of the identified genes had available microarray expression data in *S. invicta*.

## *Data availability*

Raw and filtered alignments used in these analyses, track files for the alignment editor Jalview (Clamp et al. 2004), Codeml control files and result files can be downloaded at http://bioinfo.unil.ch/supdata/Roux_positive_selection_ants/Roux_et_al_datasets.tar.gz. A simple web interface displaying gene families, Gene Ontology mapping, Codeml results and alignments (through a Jalview applet) is available at http://bioinfo.unil.ch/supdata/Roux_positive_selection_ants/families.html. Jalview tracks display the regions used or filtered out in the original protein alignments, as well as the residues found to be under positive selection by Bayes Empirical Bayes (Yang, Wong, Nielsen 2005) in all the branches tested for each of the 3 replicate runs (Figure 2).

## Acknowledgements

The authors thank Yannick Wurm, Miguel Corona Villegas, Nicolas Salamin, Corrie Moreau, and members of the Keller lab for stimulating discussions. We are grateful to Oksana Riba-Grognuz, Lino Ometto, Roberto Bonasio, Robert Waterhouse, Hannes Schabauer, Walid Gharib and members of the Ant Comparative Genomics Consortium for making data or software available for this study; Ben-Yang Liao and Meng-Pin Weng for help with Flybase phenotypic data extraction; Alexander Wild for providing illustrations for Figure 1; and four anonymous reviewers for valuable comments. Computations were performed at the Vital-IT (http://www.vital-it.ch) center for high-performance computing of the SIB Swiss Institute of Bioinformatics. JR was funded by a Swiss NSF grant attributed to LK, a Swiss NSF postdoc mobility fellowship (PBLAP3-134342) and a Marie Curie fellowship. MRR and SM acknowledge funding from a Swiss NSF grant (31003A 133011/1), the Swiss Platform for High-Performance and High-Productivity Computing (HP2C), and project UNIL.5/SMSCG as part of the AAA/SWITCH. MRR and JTD acknowledge funding from a Swiss NSF ProDoc grant (PDFMP3_130309). LK is supported by several grants from the Swiss NSF and a ERC advanced grant.

# Figure legends

Figure 1: Phylogeny of the seven sequenced ant species and the five outgroups used in this study.
The maximum likelihood phylogeny was computed by R. Waterhouse from the concatenated alignment of the conserved protein sequences of 2,756 single-copy orthologs from OrthoDB (Simola et al. 2013). The scale bar indicates the average number of amino acid substitutions per site. The phylogeny is consistent with a previously published study (Brady et al. 2006). A second study only found a difference in the branching of *P. barbatus* and *S. invicta* (Moreau et al. 2006). The 15 different branches where positive selection was tested are highlighted in red (the seven terminal branches leading to ant species and the branches numbered #1 to #8). The percentage of gene families showing positive selection in each of these branches at FDR = 10% is displayed in Table 2. Illustrations of the seven ant species and *A. mellifera* are courtesy of Alexander Wild at http://www.alexanderwild.com. *P. humanus* illustration was downloaded from Vectorbase, *D. melanogaster, T. castaneum* and *N. vitripennis* illustrations were downloaded from Wikipedia. Illustrations are not to scale.

Figure 2: Protein alignment view of positive selection signal on gene family 11650.
See Table 1 for description of the potential function of this gene family. Protein alignment is shown partially, from position 230 to 350, and *D. melanogaster*, *T. castaneum* and *P. humanus* genes were removed by MaxAlign because of insufficient alignment quality (Materials and Methods). The second annotation track under the protein alignment ("branch 5 BEB site") indicates positively selected sites on the tested branch #5 (Figure 1). Site 285 of the alignment (indicated with a red arrow) has been selected; it shows the fixation of isoleucine in lieu of the ancestral tryptophan, in the formicoid clade including 6 of the 7 ant species.

Figure 3: Positive selection in the olfactory receptors gene family. Phylogenetic tree of manually annotated protein coding sequences of olfactory receptors genes, including 291 genes from *P. barbatus* in blue, 320 genes from *L. humile* in green and 262 genes from *N. vitripennis* in black. The topology of the tree depicts the assemblage of 16 subtrees where positive selection was tested using the branch-site test of Codeml (Materials and Methods). Tested branches are depicted in gray if there was no evidence

for positive selection and in red if there was significant evidence for positive selection at 10% FDR. Untested branches are depicted in black. Scale bar indicates the number of amino acid substitutions per site.

# Tables

Table 1: Top scoring gene families at branch-site and site tests for positive selection.

Gene families are ranked based on their log-likelihood ratios (ΔlnL). Results of the branch-site test were filtered to keep only internal ant branches of the phylogenetic tree (branches #1 to #6) and with a $d_S$ on the tested branch below 1. Results of both tests were filtered to keep families with a good support for the detection of sites evolving under positive selection (BEB posterior probability > 0.9). Manual inspection of the best hits confirmed that the signal of positive selection seemed genuine for all cases, except for family 12370 in the branch-site test results, which was removed from the list.

| Test used | Gene Family | Branch | ΔlnL | p-val | FDR | $d_S$ | ω (proportion) | *D. melanogaster* gene name | Function annotated in Flybase and Uniprot | Duplicates in ants | Uniprot ID | Refs |
|---|---|---|---|---|---|---|---|---|---|---|---|---|
| **Branch-site test** | 150 | 6 | 16.1 | 1.4E-8 | 2.2E-6 | 0.93 | 281 (4.2%) | Tequila | Serine-type endopeptidase activity; long-term memory; aging | - | O45029 | (Didelot et al. 2006; Chen et al. 2012; Remolina et al. 2012) |
| | 11650[a] | 5 | 12.1 | 8.4E-7 | 7.9E-5 | 0.059 | 299 (2.5%) | CG17321 | Unknow | - | Q9VJ40 | - |
| | 453 | 6 | 11.4 | 1.9E-6 | 1.5E-4 | 0.29 | 44 (2.2%) | Guanine nucleotide exchange factor in mesoderm | Ral GTPase binding; imaginal disc-derived wing vein specification | - | A1ZBA1 | (Blanke, Jackle 2006) |
| | 361 | 1 | 10.9 | 3.0E-6 | 2.3E-4 | 0.090 | 1.2 (6.9%) | Megator | Spindle assembly | - | A1Z8P9 | (Qi et al. 2004) |
| | 5623 | 1 | 10.6 | 4.0E-6 | 3.0E-4 | 0.13 | 24 (4.6%) | Methylthioribose-1-phosphate isomerase | Catalyzes interconversion of methylthioribose-1-phosphate into methylthioribulose-1-phosphate; wing disc development | - | Q9V9X4 | (Bronstein et al. 2010) |
| | 1050 | 6 | 10.1 | 6.7E-6 | 4.7E-4 | 0.49 | 45 (3%) | Dis3 | Regulation of gene expression; nuclear RNA surveillance; neurogenesis | - | Q8MSY2 | (Kuan et al. 2009; Kiss, Andrulis 2010; Neumuller et al. 2011) |
| | 793 | 4 | 9.6 | 1.2E-5 | 7.6E-4 | 0.085 | ∞ (0.6%) | embargoed | Protein binding; protein transporter activity; protein export from nucleus; multicellular organismal development; centriole replication | - | Q9TVM2 | (Collier et al. 2000; Roth et al. 2003) |
| | 8639 | 6 | 9.5 | 1.4E-5 | 8.3E-4 | 0.26 | ∞ (11%) | ATP synthase, subunit b, mitochondria | Hydrogen-exporting ATPase activity, phosphorylative mechanism; phagocytosis, engulfment | - | Q94516 | (Stroschein-Stevenson et al. 2005) |
| | 3983 | 4 | 8.9 | 2.4E-5 | 0.0014 | 0.036 | ∞ (0.5%) | lysyl oxidase-like 2 | Protein-lysine 6-oxidase activity | - | Q8IH65 | (Molnar et al. 2003; Molnar et al. 2005) |
| | 2208 | 6 | 8.7 | 3.0E-5 | 0.0016 | 0.49 | ∞ (1.2%) | Cytochrome P450 reductase | NADPH-hemoprotein reductase activity; oxidation-reduction process; putative function in olfactory clearance | - | Q27597 | (Hovemann, Sehlmeyer, Malz 1997) |
| **Site test** | 3245 | - | 10.6 | 4.2E-6 | 0.0041 | - | 2.4 (2.1%) | CG6752 | Unknown | Yes | Q9VFC4, Q8SZS1 | - |
| | 6214 | - | 8.3 | 4.6E-5 | 0.038 | - | 8.3 (0.9%) | CG42343 | Unknown | No | B7Z153, Q9VRI6 | - |
| | 6649 | - | 8.0 | 6.6E-5 | 0.045 | - | 4.9 (4.7%) | CG7845 | Muscle cell homeostasis | No | Q7K4B2 | (Kucherenko et al. 2011) |
| | 5707 | - | 7.9 | 7.2E-5 | 0.045 | - | 3.4 (2.1%) | mitochondrial ribosomal protein L37 | Structural constituent of ribosome; translation | No | Q9VGW9, Q3YNF4, Q3YNF5 | (Kim et al. 2010a) |
| | 2372 | - | 7.8 | 7.6E-5 | 0.045 | - | 3.2 (1.9%) | Mitochondrial trifunctional protein α subunit | Long-chain-3-hydroxyacyl-CoA dehydrogenase activity; long-chain-enoyl-CoA hydratase activity; response to starvation; determination of adult lifespan; fatty acid beta-oxidation; wound healing | No | Q8IPE8, Q9V397 | (Kishita, Tsuda, Aigaki 2012) |
| | 8490 | - | 7.4 | 1.2E-4 | 0.062 | - | 6.2 (2.0%) | Phosphatidylinositol synthase | CDP-diacylglycerol-inositol 3-phosphatidyltransferase activity; phototransduction | No | Q8IR29, Q8SX37 | (Wang, Montell 2006) |
| | 3891 | - | 6.8 | 2.2E-4 | 0.11 | - | 9.4 (0.3%) | CG1607 | Potential amino acid transmembrane transporter activity | No | Q9V9Y0, Q95T33 | - |
| | 2074 | - | 6.5 | 3.1E-4 | 0.14 | - | 4.2 (4.0%) | CG9715 | Unknown | Yes | Q9VVA9, Q960D5 | - |
| | 1584 | - | 6.3 | 4.0E-4 | 0.17 | - | 2.3 (2.7%) | unextended | Potential role in cellular ion homeostasis | No | A8Y516 | - |
| | 1053 | - | 6.2 | 4.2E-4 | 0.17 | - | 9.1 (0.2%) | Coat Protein (coatomer) β | Biosynthetic protein transport from the ER, via the Golgi up to the trans Golgi network. Required for limiting lipid storage in lipid droplets. Involved in innate immune response. | No | P45437 | (Bard et al. 2006; Beller et al. 2008; Cronin et al. 2009) |

[a] Example used in Figure 2

Table 2: Amount of positive selection detected on different branches of the analyzed phylogeny

| **Branch name[a]** | **Lineage delineated** | **Fraction of positively selected gene families** | **Number of positively selected gene families[b]** |
|---|---|---|---|
| Acep | *A. cephalotes* | 0.056 | 144 |
| Aech | *A. echinatior* | 0.043 | 109 |
| Sinv | *S. invicta* | 0.029 | 85 |
| Pbar | *P. barbatus* | 0.038 | 80 |
| Cflo | *C. floridae* | 0.017 | 65 |
| Lhum | *L. humile* | 0.036 | 97 |
| Hsal | *H. saltator* | 0.020 | 76 |
| 1 | Attini | 0.0088 | 16 |
| 2 | Myrmicinae | 0.0071 | 10 |
| 3 | Myrmicinae | 0.0087 | 16 |
| 4 | Formicoid | 0.0072 | 17 |
| 5 | Formicoid | 0.025 | 58 |
| 6 | Formicidae | 0.030 | 87 |
| 7 | Aculeata | 0.10 | 176 |
| 8 | Hymenoptera/Apocrita | 0.39 | 762 |
| All above branches except 7 and 8 | Formicidae | 0.20 | 830 |
| All above branches | Hymenoptera/Apocrita | 0.43 | 1,832 |

[a] As illustrated in Figure 1

[b] Branches of gene families trees can be merged if genes are missing (or removed for quality reasons), and the resulting branches do not correspond to canonical branches defined by the species topology (Figure 1). When positive selection is found on such branches, it was not counted in branch-specific numbers displayed in Table 2, but it was counted when a whole lineage was considered (e.g., Hymenoptera).

Table 3: Gene Ontology categories enriched for positively selected genes, based on scores from the branch-site test from Codeml in ants. The enrichment test considers a combined score for all analyzed branches of the ant lineage (Materials and Methods). The full table of results is shown in Table S6.

| setID | Ontology | setName | setSize | score | p-value | FDR |
|---|---|---|---|---|---|---|
| GO:0000313 | CC | organellar ribosome | 59 | 26.8 | 1.4E-10 | 0 |
| GO:0006120 | BP | mitochondrial electron transport, NADH to ubiquinone | 18 | 10.6 | 1.1E-9 | 0 |
| GO:0005759 | CC | mitochondrial matrix | 98 | 39.7 | 1.6E-9 | 0 |
| GO:0005762 | CC | mitochondrial large ribosomal subunit | 36 | 16.8 | 1.1E-7 | 0.0025 |
| GO:0005746 | CC | mitochondrial respiratory chain | 31 | 14.6 | 4.5E-7 | 0.0033 |
| GO:0005747 | CC | mitochondrial respiratory chain complex I | 22 | 11.0 | 1.3E-6 | 0.0033 |
| GO:0008137 | MF | NADH dehydrogenase (ubiquinone) activity | 16 | 8.0 | 3.2E-5 | 0.013 |
| GO:0005763 | CC | mitochondrial small ribosomal subunit | 25 | 10.9 | 0.00018 | 0.047 |
| GO:0008038 | BP | neuron recognition | 19 | 8.7 | 0.00023 | 0.047 |
| GO:0008344 | BP | adult locomotory behavior | 19 | 8.4 | 0.00082 | 0.086 |
| GO:0042254 | BP | ribosome biogenesis | 39 | 15.0 | 0.0011 | 0.099 |
| GO:0003735 | MF | structural constituent of ribosome | 107 | 36.4 | 0.0012 | 0.099 |
| GO:0044459 | CC | plasma membrane part | 129 | 42.9 | 0.0016 | 0.12 |
| GO:0006508 | BP | Proteolysis | 145 | 47.4 | 0.0022 | 0.14 |
| GO:0006412 | BP | Translation | 191 | 61.0 | 0.0025 | 0.15 |
| GO:0016491 | MF | oxidoreductase activity | 127 | 41.8 | 0.0028 | 0.15 |
| GO:0004872 | MF | receptor activity | 90 | 30.6 | 0.0028 | 0.15 |
| GO:0055114 | BP | oxidation-reduction process | 129 | 42.2 | 0.0038 | 0.16 |
| GO:0008237 | MF | metallopeptidase activity | 36 | 13.6 | 0.0039 | 0.16 |
| GO:0061134 | MF | peptidase regulator activity | 17 | 7.2 | 0.0046 | 0.18 |
| GO:0002520 | BP | immune system development | 26 | 10.2 | 0.0053 | 0.19 |
| GO:0048534 | BP | hemopoietic or lymphoid organ development | 26 | 10.2 | 0.0053 | 0.19 |
| GO:0016616 | MF | oxidoreductase activity, acting on the CH-OH group of donors, NAD or NADP as acceptor | 18 | 7.5 | 0.0053 | 0.19 |
| GO:0016836 | MF | hydro-lyase activity | 14 | 6.1 | 0.0055 | 0.19 |

Table 4: Gene Ontology categories enriched for positively selected genes, based on scores from the branch-site test from Codeml in *Drosophila*. Depletion results are shown in Table S4.

| setID | Ontology | setName | setSize | score | p-value | FDR |
|---|---|---|---|---|---|---|
| GO:0006030 | BP | chitin metabolic process | 29 | 16.2 | 4.0E-6 | 0.0015 |
| GO:0006022 | BP | aminoglycan metabolic process | 36 | 19.4 | 5.7E-6 | 0.0018 |
| GO:0006952 | BP | defense response | 36 | 19.2 | 9.8E-6 | 0.0018 |
| GO:0008061 | MF | chitin binding | 24 | 13.5 | 2.0E-5 | 0.0020 |
| GO:0004252 | MF | serine-type endopeptidase activity | 52 | 26.0 | 3.3E-5 | 0.0023 |
| GO:0008026 | MF | ATP-dependent helicase activity | 18 | 10.5 | 4.0E-5 | 0.0026 |
| GO:0004872 | MF | receptor activity | 13 | 7.9 | 8.5E-5 | 0.0048 |
| GO:0006006 | BP | glucose metabolic process | 13 | 7.8 | 0.00021 | 0.0082 |
| GO:0046486 | BP | glycerolipid metabolic process | 17 | 9.7 | 0.00023 | 0.0090 |
| GO:0005819 | CC | spindle | 20 | 10.9 | 0.00046 | 0.012 |
| GO:0004175 | MF | endopeptidase activity | 78 | 35.9 | 0.00048 | 0.012 |
| GO:0009607 | BP | response to biotic stimulus | 31 | 15.8 | 0.00060 | 0.013 |
| GO:0051707 | BP | response to other organism | 31 | 15.8 | 0.00060 | 0.013 |
| GO:0006508 | BP | proteolysis | 136 | 59.5 | 0.00071 | 0.014 |
| GO:0006007 | BP | glucose catabolic process | 12 | 7.0 | 0.00072 | 0.014 |
| GO:0019320 | BP | hexose catabolic process | 12 | 7.0 | 0.00072 | 0.014 |
| GO:0030312 | CC | external encapsulating structure | 12 | 7.0 | 0.00074 | 0.014 |
| GO:0015081 | MF | sodium ion transmembrane transporter activity | 16 | 8.9 | 0.00088 | 0.016 |
| GO:0051649 | BP | establishment of localization in cell | 14 | 7.9 | 0.00093 | 0.017 |
| GO:0007126 | BP | meiosis | 34 | 16.9 | 0.00096 | 0.017 |
| GO:0003824 | MF | catalytic activity | 844 | 337.0 | 0.0011 | 0.018 |
| GO:0046488 | BP | phosphatidylinositol metabolic process | 11 | 6.5 | 0.0012 | 0.018 |
| GO:0016490 | MF | structural constituent of peritrophic membrane | 11 | 6.4 | 0.0013 | 0.018 |
| GO:0005975 | BP | carbohydrate metabolic process | 64 | 29.5 | 0.0015 | 0.019 |
| GO:0004888 | MF | transmembrane receptor activity | 49 | 23.2 | 0.0016 | 0.020 |
| GO:0051276 | BP | chromosome organization | 59 | 27.2 | 0.0020 | 0.024 |
| GO:0008270 | MF | zinc ion binding | 173 | 73.4 | 0.0027 | 0.029 |
| GO:0002376 | BP | immune system process | 43 | 20.4 | 0.0027 | 0.029 |
| GO:0002759 | BP | regulation of antimicrobial humoral response | 11 | 6.3 | 0.0028 | 0.029 |
| GO:0004984 | MF | olfactory receptor activity | 19 | 9.9 | 0.0030 | 0.029 |
| GO:0002697 | BP | regulation of immune effector process | 11 | 6.2 | 0.0038 | 0.035 |
| GO:0000819 | BP | sister chromatid segregation | 11 | 6.2 | 0.0038 | 0.035 |
| GO:0007143 | BP | female meiosis | 11 | 6.2 | 0.0047 | 0.040 |
| GO:0016021 | CC | integral to membrane | 224 | 93.1 | 0.0047 | 0.040 |
| GO:0031347 | BP | regulation of defense response | 12 | 6.6 | 0.0055 | 0.044 |
| GO:0015370 | MF | solute:sodium symporter activity | 12 | 6.6 | 0.0061 | 0.047 |
| GO:0000272 | BP | polysaccharide catabolic process | 11 | 6.1 | 0.0065 | 0.049 |
| GO:0016810 | MF | hydrolase activity, acting on carbon-nitrogen (but not peptide) bonds | 28 | 13.6 | 0.0065 | 0.049 |
| GO:0004521 | MF | endoribonuclease activity | 11 | 6.1 | 0.0069 | 0.052 |
| GO:0007291 | BP | sperm individualization | 14 | 7.4 | 0.0074 | 0.053 |
| GO:0010564 | BP | regulation of cell cycle process | 30 | 14.4 | 0.0075 | 0.053 |
| GO:0005635 | CC | nuclear envelope | 17 | 8.7 | 0.0077 | 0.054 |
| GO:0016773 | MF | phosphotransferase activity, alcohol group as acceptor | 82 | 36.0 | 0.0077 | 0.054 |
| GO:0051253 | BP | negative regulation of RNA metabolic process | 35 | 16.5 | 0.0080 | 0.055 |
| GO:0007608 | BP | sensory perception of smell | 21 | 10.5 | 0.0080 | 0.055 |
| GO:0004222 | MF | metalloendopeptidase activity | 26 | 12.7 | 0.0082 | 0.055 |
| GO:0006807 | BP | nitrogen compound metabolic process | 298 | 121.5 | 0.0090 | 0.059 |
| GO:0005576 | CC | extracellular region | 97 | 41.9 | 0.010 | 0.066 |
| GO:0006814 | BP | sodium ion transport | 19 | 9.5 | 0.011 | 0.068 |
| GO:0045132 | BP | meiotic chromosome segregation | 11 | 5.9 | 0.011 | 0.070 |
| GO:0034641 | BP | cellular nitrogen compound metabolic process | 296 | 120.4 | 0.012 | 0.072 |
| GO:0010629 | BP | negative regulation of gene expression | 42 | 19.2 | 0.013 | 0.073 |
| GO:0090304 | BP | nucleic acid metabolic process | 162 | 67.6 | 0.013 | 0.075 |
| GO:0016301 | MF | kinase activity | 91 | 39.2 | 0.013 | 0.075 |
| GO:0048584 | BP | positive regulation of response to stimulus | 11 | 5.9 | 0.015 | 0.084 |
| GO:0016798 | MF | hydrolase activity, acting on glycosyl bonds | 26 | 12.4 | 0.017 | 0.086 |
| GO:0006139 | BP | nucleobase, nucleoside, nucleotide and nucleic acid metabolic process | 236 | 96.4 | 0.017 | 0.086 |
| GO:0016491 | MF | oxidoreductase activity | 201 | 82.6 | 0.018 | 0.090 |
| GO:0009987 | BP | cellular process | 790 | 310.5 | 0.019 | 0.095 |
| GO:0007088 | BP | regulation of mitosis | 16 | 8.0 | 0.019 | 0.095 |
| GO:0051783 | BP | regulation of nuclear division | 16 | 8.0 | 0.019 | 0.095 |
| GO:0006810 | BP | transport | 200 | 82.1 | 0.021 | 0.10 |
| GO:0051234 | BP | establishment of localization | 197 | 80.9 | 0.021 | 0.10 |
| GO:0006066 | BP | alcohol metabolic process | 35 | 16.1 | 0.021 | 0.10 |
| GO:0004553 | MF | hydrolase activity, hydrolyzing O-glycosyl compounds | 22 | 10.6 | 0.022 | 0.11 |
| GO:0008233 | MF | peptidase activity | 26 | 12.3 | 0.022 | 0.11 |
| GO:0070011 | MF | peptidase activity, acting on L-amino acid peptides | 22 | 10.6 | 0.022 | 0.11 |
| GO:0046914 | MF | transition metal ion binding | 58 | 25.5 | 0.022 | 0.11 |
| GO:0050660 | MF | flavin adenine dinucleotide binding | 16 | 8.0 | 0.024 | 0.11 |
| GO:0045892 | BP | negative regulation of transcription, DNA-dependent | 26 | 12.2 | 0.024 | 0.11 |
| GO:0032553 | MF | ribonucleotide binding | 161 | 66.5 | 0.024 | 0.11 |
| GO:0032555 | MF | purine ribonucleotide binding | 161 | 66.5 | 0.024 | 0.11 |
| GO:0035639 | MF | purine ribonucleoside triphosphate binding | 161 | 66.5 | 0.024 | 0.11 |
| GO:0006396 | BP | RNA processing | 36 | 16.4 | 0.025 | 0.11 |
| GO:0031226 | CC | intrinsic to plasma membrane | 30 | 13.9 | 0.026 | 0.11 |
| GO:0035222 | BP | wing disc pattern formation | 11 | 5.7 | 0.026 | 0.12 |
| GO:0007346 | BP | regulation of mitotic cell cycle | 31 | 14.3 | 0.028 | 0.12 |
| GO:0045017 | BP | glycerolipid biosynthetic process | 11 | 5.7 | 0.030 | 0.13 |
| GO:0006955 | BP | immune response | 30 | 13.8 | 0.030 | 0.13 |
| GO:0044262 | BP | cellular carbohydrate metabolic process | 44 | 19.6 | 0.030 | 0.13 |
| GO:0017076 | MF | purine nucleotide binding | 164 | 67.5 | 0.032 | 0.14 |
| GO:0016705 | MF | oxidoreductase activity, acting on paired donors, with incorporation or reduction of molecular oxygen | 21 | 10.0 | 0.032 | 0.14 |
| GO:0005524 | MF | ATP binding | 163 | 67.0 | 0.034 | 0.14 |
| GO:0030554 | MF | adenyl nucleotide binding | 163 | 67.0 | 0.034 | 0.14 |
| GO:0032559 | MF | adenyl ribonucleotide binding | 163 | 67.0 | 0.034 | 0.14 |
| GO:0008237 | MF | metallopeptidase activity | 12 | 6.1 | 0.034 | 0.14 |
| GO:0007127 | BP | meiosis I | 18 | 8.7 | 0.035 | 0.14 |
| GO:0019730 | BP | antimicrobial humoral response | 14 | 6.9 | 0.039 | 0.15 |
| GO:0005815 | CC | microtubule organizing center | 16 | 7.8 | 0.041 | 0.16 |
| GO:0055114 | BP | oxidation-reduction process | 167 | 68.3 | 0.043 | 0.17 |
| GO:0019899 | MF | enzyme binding | 14 | 6.9 | 0.044 | 0.17 |
| GO:0048232 | BP | male gamete generation | 45 | 19.8 | 0.045 | 0.17 |
| GO:0008033 | BP | tRNA processing | 17 | 8.2 | 0.045 | 0.17 |
| GO:0005887 | CC | integral to plasma membrane | 29 | 13.2 | 0.046 | 0.17 |
| GO:0044281 | BP | small molecule metabolic process | 212 | 85.8 | 0.046 | 0.17 |
| GO:0008238 | MF | exopeptidase activity | 18 | 8.6 | 0.046 | 0.17 |
| GO:0051179 | BP | localization | 236 | 95.1 | 0.047 | 0.17 |
| GO:0007283 | BP | spermatogenesis | 44 | 19.3 | 0.047 | 0.18 |
| GO:0050662 | MF | coenzyme binding | 48 | 21.0 | 0.048 | 0.18 |
| GO:0034470 | BP | ncRNA processing | 27 | 12.3 | 0.049 | 0.18 |
| GO:0048515 | BP | spermatid differentiation | 24 | 11.1 | 0.050 | 0.18 |
| GO:0045786 | BP | negative regulation of cell cycle | 11 | 5.5 | 0.050 | 0.18 |
| GO:0045934 | BP | negative regulation of nucleobase, nucleoside, nucleotide and nucleic acid metabolic process | 41 | 18.1 | 0.052 | 0.18 |
| GO:0010639 | BP | negative regulation of organelle organization | 12 | 5.9 | 0.055 | 0.19 |
| GO:0015631 | MF | tubulin binding | 12 | 5.9 | 0.056 | 0.19 |
| GO:0005549 | MF | odorant binding | 43 | 18.8 | 0.058 | 0.20 |

Table 5: Gene Ontology categories enriched for positively selected genes, based on scores from the branch-site test from Codeml in bees. Depletion results are shown in Table S5.

| **setID** | **Ontology** | **setName** | **setSize** | **score** | **p-value** | **FDR** |
|---|---|---|---|---|---|---|
| GO:0005099 | MF | Ras GTPase activator activity | 11 | 6.0 | 1.1E-5 | 0.03 |
| GO:0005083 | MF | small GTPase regulator activity | 18 | 8.1 | 0.00010 | 0.041 |
| GO:0004872 | MF | receptor activity | 16 | 7.2 | 0.00020 | 0.041 |
| GO:0022836 | MF | gated channel activity | 11 | 5.4 | 0.00021 | 0.041 |
| GO:0006399 | BP | tRNA metabolic process | 26 | 10.3 | 0.00040 | 0.053 |
| GO:0071842 | BP | cellular component organization at cellular level | 190 | 55.5 | 0.0013 | 0.065 |
| GO:0006418 | BP | tRNA aminoacylation for protein translation | 19 | 7.6 | 0.0021 | 0.084 |
| GO:0009725 | BP | response to hormone stimulus | 11 | 4.9 | 0.0022 | 0.084 |
| GO:0005635 | CC | nuclear envelope | 13 | 5.6 | 0.0022 | 0.084 |
| GO:0032507 | BP | maintenance of protein location in cell | 12 | 5.3 | 0.0023 | 0.084 |
| GO:0051336 | BP | regulation of hydrolase activity | 20 | 7.9 | 0.0026 | 0.089 |
| GO:0006629 | BP | lipid metabolic process | 51 | 17.0 | 0.0031 | 0.095 |
| GO:0031072 | MF | heat shock protein binding | 17 | 6.7 | 0.0046 | 0.11 |
| GO:0008152 | BP | metabolic process | 335 | 91.3 | 0.0070 | 0.14 |
| GO:0004812 | MF | aminoacyl-tRNA ligase activity | 19 | 7.2 | 0.0075 | 0.14 |
| GO:0016740 | MF | transferase activity | 211 | 59.1 | 0.0087 | 0.14 |
| GO:0019899 | MF | enzyme binding | 19 | 7.1 | 0.0097 | 0.14 |
| GO:0005216 | MF | ion channel activity | 14 | 5.5 | 0.011 | 0.14 |
| GO:0022838 | MF | substrate-specific channel activity | 14 | 5.5 | 0.011 | 0.14 |
| GO:0009308 | BP | amine metabolic process | 47 | 15.2 | 0.011 | 0.14 |
| GO:0004222 | MF | metalloendopeptidase activity | 15 | 5.8 | 0.012 | 0.14 |
| GO:0005938 | CC | cell cortex | 15 | 5.8 | 0.012 | 0.14 |
| GO:0008237 | MF | metallopeptidase activity | 23 | 8.2 | 0.012 | 0.14 |
| GO:0007275 | BP | multicellular organismal development | 274 | 74.9 | 0.013 | 0.14 |
| GO:0007165 | BP | signal transduction | 97 | 28.8 | 0.013 | 0.14 |
| GO:0044106 | BP | cellular amine metabolic process | 40 | 12.9 | 0.019 | 0.19 |
| GO:0044459 | CC | plasma membrane part | 45 | 14.3 | 0.021 | 0.19 |
| GO:0006520 | BP | cellular amino acid metabolic process | 32 | 10.6 | 0.021 | 0.19 |
| GO:0032879 | BP | regulation of localization | 36 | 11.7 | 0.022 | 0.19 |
| GO:0006140 | BP | regulation of nucleotide metabolic process | 13 | 5.0 | 0.022 | 0.19 |
| GO:0030811 | BP | regulation of nucleotide catabolic process | 13 | 5.0 | 0.022 | 0.19 |
| GO:0033121 | BP | regulation of purine nucleotide catabolic process | 13 | 5.0 | 0.022 | 0.19 |
| GO:0033124 | BP | regulation of GTP catabolic process | 13 | 5.0 | 0.022 | 0.19 |
| GO:0043087 | BP | regulation of GTPase activity | 13 | 5.0 | 0.022 | 0.19 |
| GO:0006793 | BP | phosphorus metabolic process | 97 | 28.3 | 0.023 | 0.19 |
| GO:0006796 | BP | phosphate metabolic process | 97 | 28.3 | 0.023 | 0.19 |
| GO:0042578 | MF | phosphoric ester hydrolase activity | 42 | 13.4 | 0.024 | 0.19 |
| GO:0016758 | MF | transferase activity, transferring hexosyl groups | 26 | 8.8 | 0.024 | 0.19 |

# Supporting information

## *Supplementary text, methods, figures and references in separate PDF file.*

## *Supplementary tables in separate Excel files:*

Table S1: Results from the branch-site test for positive selection applied on the ant single-copy orthologs dataset (4,256 families). Family numbering as in OrthoDB run (http://bioinfo.unil.ch/supdata/Roux_positive_selection_ants/orthoDB_run.zip). Branch numbering as in Figure 1. A "comparable" branch means that in this particular gene family, the topology of this particular branch is identical to the species tree, and not affected by missing genes.

Table S2: Results from the branch-site test for positive selection applied on the bee dataset (2,256 families). Family numbering based on *Apis mellifera* orthologs. Branch numbering as in Figure S5. Legend as Table S1.

Table S3: Results from the branch-site test for positive selection applied on the twelve Drosophila dataset (3,749 families). Family numbering based on *Drosophila melanogaster* orthologs. Branch numbering as in Figure S4. Legend as Table S1.

Table S4: SUMSTAT gene set enrichment test on Gene Ontology functional categories based on scores for positive selection in 12 Drosophila using the branch-site test from Codeml (combined score for all analyzed branches of the Drosophila lineage).

Table S5: SUMSTAT gene set enrichment test on Gene Ontology functional categories based on scores for positive selection in bees using the branch-site test from Codeml (combined for all analyzed branches of the bee lineage, with the exception of two long basal branches leading to *Megachile rotundata* and *Exoneura robusta*).

Table S6: Gene Ontology enrichment test based on scores from the branch-site test from Codeml in ants. This is the full table of all GO categories tested. See Table 3 for only the GO categories that are significantly enriched or depleted.

Table S7: Biases in Gene Ontology functional categories for genes present in the dataset of bee single-copy orthologs gene families (10 species).

Table S8: Biases in Gene Ontology functional categories for genes present in the dataset of single-copy orthologs gene families (7 ant species and 12 species in total).

Table S9: SUMSTAT gene set enrichment test on Gene Ontology functional categories based on scores for positive selection using the branch-site test from Codeml (an enrichment test is run for each branch separately).

Table S10: Biases in Gene Ontology functional categories for genes present in the extensive ant dataset including gene families that experienced duplications (7 ant species and 12 species in total).

Table S11: Biases in Gene Ontology functional categories for genes present in the dataset of Drosophila single-copy orthologs gene families (12 species).

Table S12: SUMSTAT gene set enrichment test on phenotypic categories based on positive selection scores using the branch-site test from Codeml (an enrichment test is run for each branch separately).

Table S13: SUMSTAT gene set enrichment test on Gene Ontology functional categories based on scores for positive selection in ants using the site test from Codeml.

Table S14: Results from the site test for positive selection applied on the mitochondrial genome dataset (13 families).

Table S15: Results from the branch-site test for positive selection applied on the mitochondrial genome dataset (13 families). Branch numbering as in Figure S3. Legend as Table S1.

Table S16: SUMSTAT gene set enrichment test on sets of aging and oxidative stress differentially expressed genes in *D. melanogaster*, based on scores for positive selection using the branch-site test from Codeml (combined scores for all analyzed branches of the ant lineage).

Table S17: SUMSTAT gene set enrichment test on sets of caste-specific differentially expressed genes in *S. invicta*, based on scores for positive selection using the branch-site test from Codeml (combined scores for all analyzed branches of the ant lineage).

Table S18: Results from the branch-site test for positive selection applied on the olfactory receptors dataset. Clades and branches IDs are similar to Newick trees show in Supplementary text. Full phylogenetic tree is displayed as Figure 3 and Figure S1.

Table S19: Results from the site test for positive selection applied on the extensive ant dataset (6,186 families). Family numbering as in OrthoDB run (http://bioinfo.unil.ch/supdata/Roux_positive_selection_ants/orthoDB_run.zip).

Table S20: SUMSTAT gene set enrichment test on Gene Ontology functional categories based on scores for positive selection in ants using the branch-site test from Codeml

(combined score for all analyzed branches of the ant lineage). Only results on internal branches were considered (no leaves).

Table S21: SUMSTAT gene set enrichment test on Gene Ontology functional categories based on scores for positive selection in ants using the branch-site test from Codeml (combined score for all analyzed branches of the ant lineage). We considered only results on branches were the rate of synonymous substitutions $d_S$ was lower than 1.

Table S22: SUMSTAT gene set enrichment test on Gene Ontology functional categories based on scores for positive selection in ants using the branch-site test from Codeml (combined score for all analyzed branches of the ant lineage). Only positive results with identified sites on the alignment (BEB posterior probability > 0.9) were considered.

Table S23: SUMSTAT gene set enrichment test on Gene Ontology functional categories based on scores for positive selection in ants using the branch-site test from Codeml (combined score for all analyzed branches of the ant lineage). Only results where the 3 independent runs of Codeml did not display convergence issues were considered.

Table S24: SUMSTAT gene set enrichment test on Gene Ontology functional categories based on scores for positive selection in ants using the branch-site test from Codeml (combined score for all analyzed branches of the ant lineage). Only branches where the increase in G+C content at third codon positions was lower than 10% were considered.

Table S25: SUMSTAT gene set enrichment test on Gene Ontology functional categories based on scores for positive selection in ants using the branch-site test from Codeml (combined score for all analyzed branches of the ant lineage). Codeml was run on PAGAN multiple sequence alignments filtered with GUIDANCE.

Table S26: SUMSTAT gene set enrichment test on KEGG pathways, based on scores for positive selection in ants using the branch-site test from Codeml (combined score for all analyzed branches of the ant lineage).

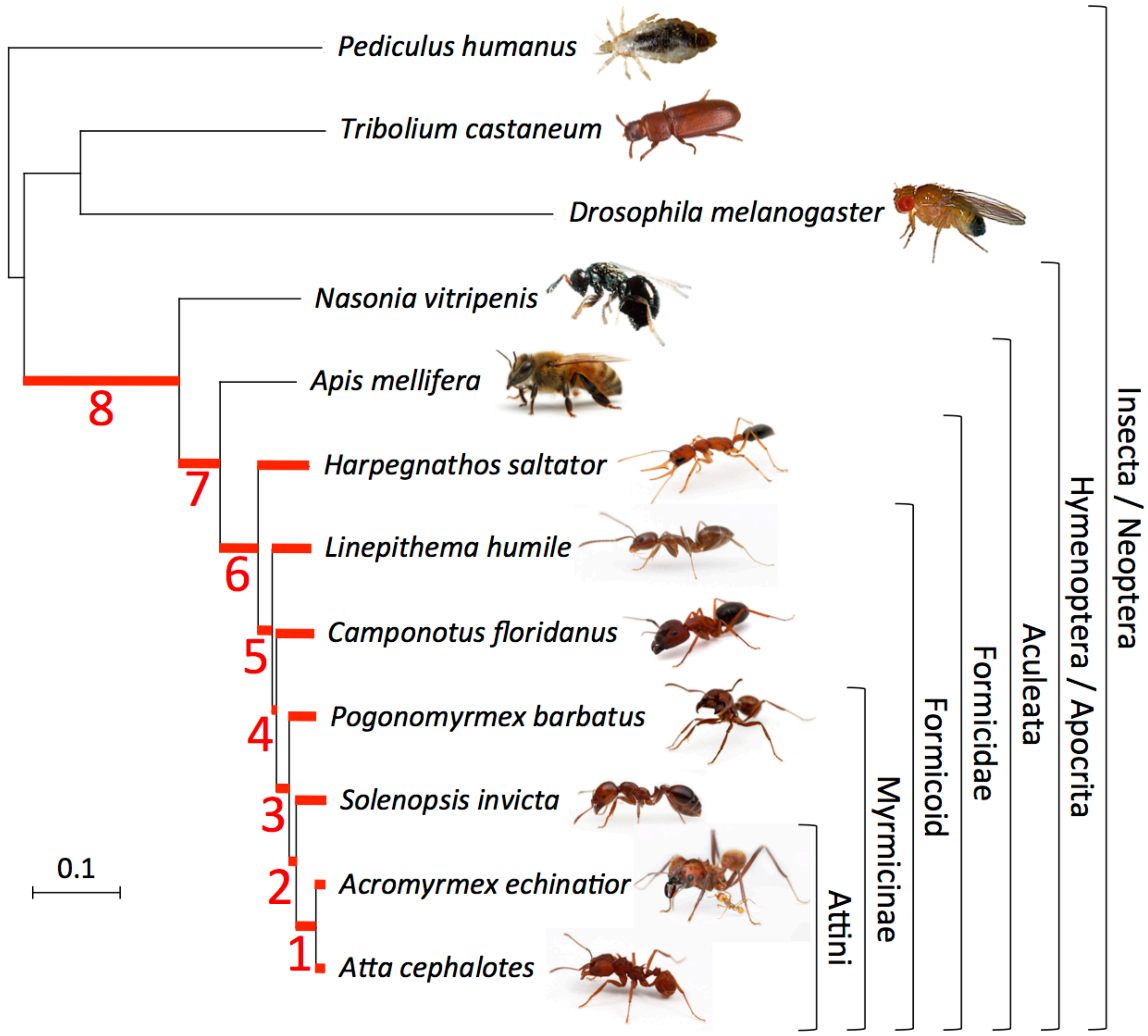

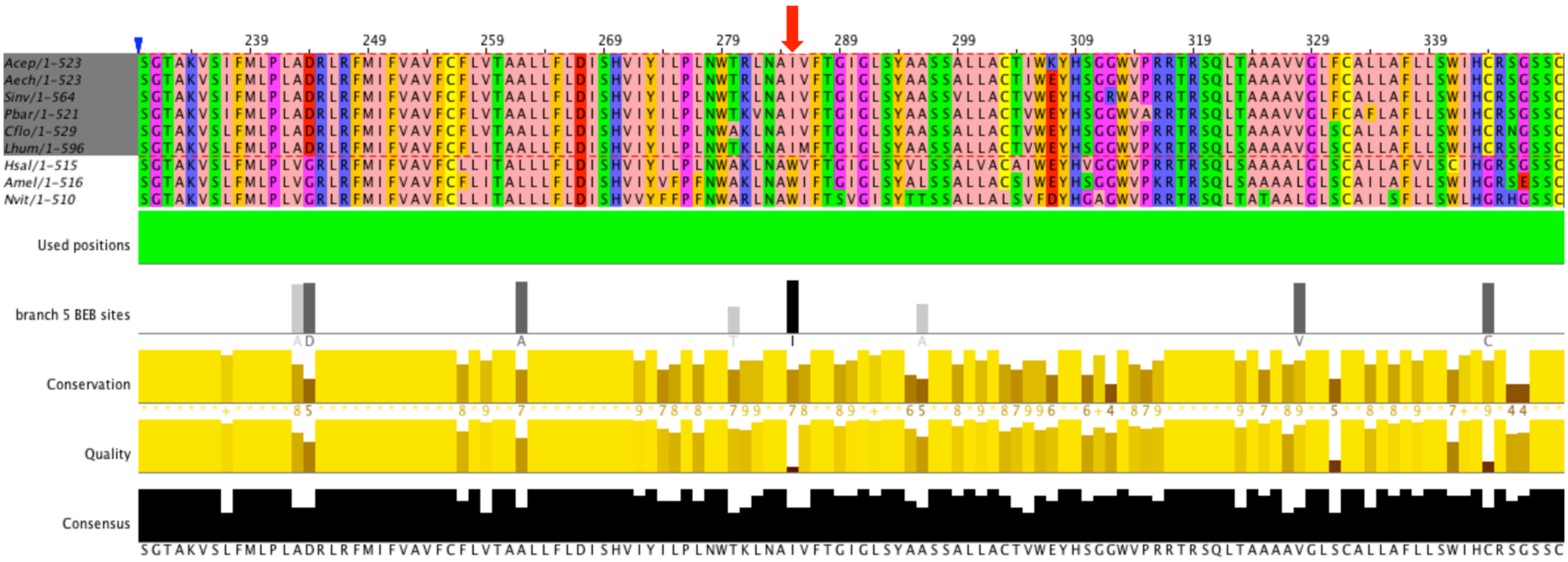

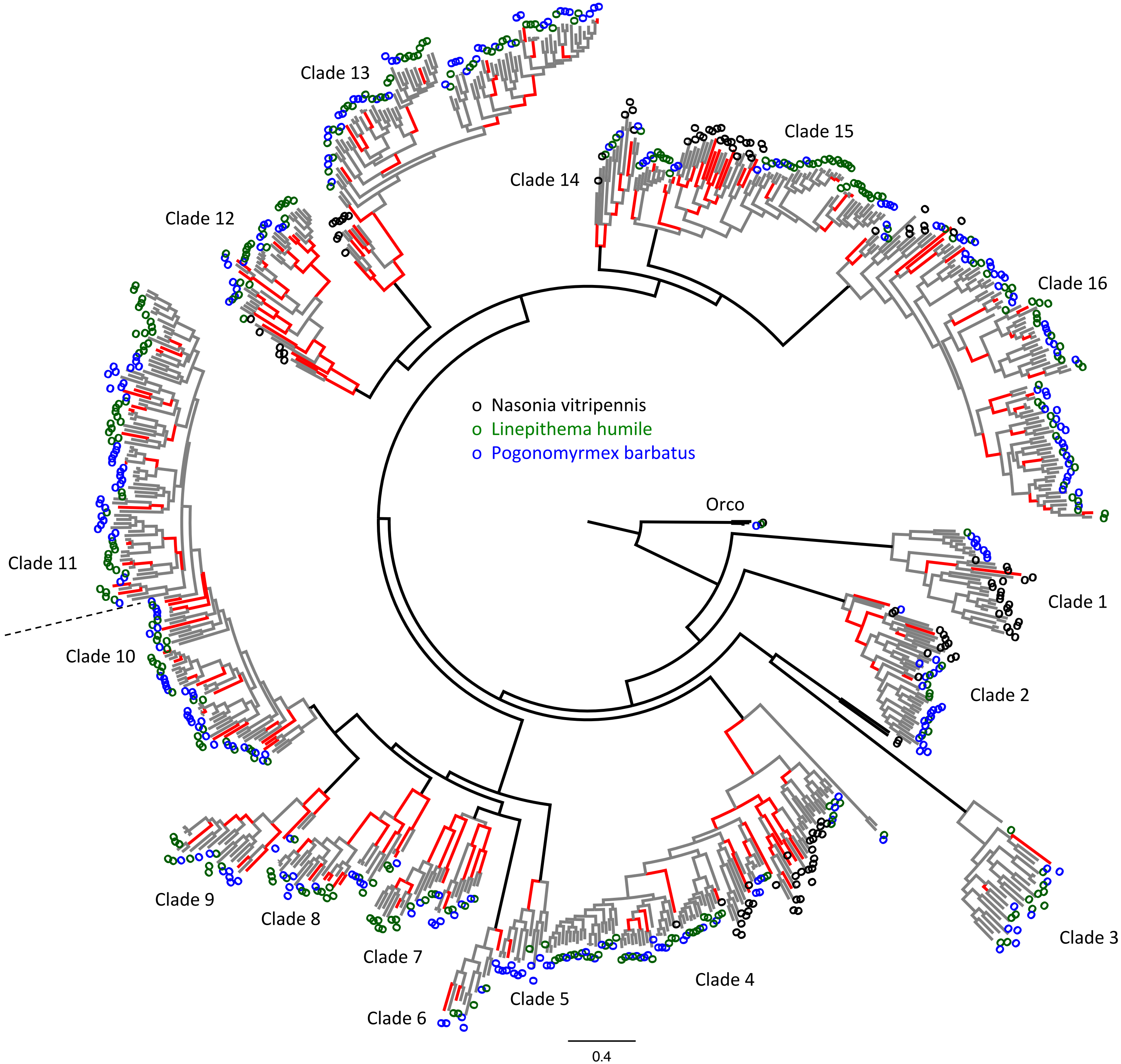